\documentclass[a4paper,11pt]{article}
\pdfoutput=1 

\usepackage{jcappub} 

\usepackage[T1]{fontenc} 

\usepackage[T1]{fontenc}
\usepackage{ae,aecompl}

\newcommand{\mo}{\mathop}
\newcommand{\us}{\mskip3mu}
\usepackage{subfig}
\usepackage{caption}
\usepackage{epigraph}
\usepackage{amsmath,amssymb}

\title{The Two Body Problem in the Presence of Dark Energy and Modified Gravity: Application to the Local Group}

\author[a]{M. McLeod}
\author[a]{O. Lahav.}

\affiliation[a]{Department of Physics and Astronomy, University College London, Gower Street, London, WC1E 6BT, UK}

\emailAdd{michael.mcleod.13@ucl.ac.uk}
\emailAdd{o.lahav@ucl.ac.uk}

\abstract{We explore mass estimation of the Local Group via the use of the simple, dynamical `timing argument' in the context of a variety of theories of dark energy and modified gravity: a cosmological constant, a perfect fluid with constant equation of state $w$, quintessence (minimally coupled scalar field), MOND, and symmetrons (coupled scalar field). We explore generic coupled scalar field theories, with the symmetron model as an explicit example. We find that theories which attempt to eliminate dark matter by fitting rotation curves produce mass estimates in the timing argument which are not compatible with the luminous mass of the galaxies alone. Assuming that the galaxies are approaching their first encounter, MOND gives a mass of around $2.7\times 10^{10} M_\odot$, roughly 10\% of the luminous mass of the LG, although a higher mass can be obtained in the case of a previous fly-by event between the MW and M31. The symmetron model suggests a mass too high to be explained without additional dark matter ($\mathcal{O}(10^{12}) M_\odot$), suggesting that there is a missing mass problem in this model. 
We also demonstrate that tensions in measurements of $H_0$ can produce an uncertainty in the Local Group mass estimate comparable to observational uncertainties on the separation and relative velocity of the galaxies, with values for the mass ranging from $4.5 - 5.4 \times 10^{12} M_{\odot}$ varying $h$ between 0.67 and 0.76. }

\begin{document}
\maketitle
\flushbottom

\section{Introduction}\label{intro_3}

The `timing argument' (TA) has been widely used in the history of the literature concerning the mass of the Local Group (LG), originating in a landmark paper in 1959 \cite{KW}; nevertheless it remains an outstanding issue that dynamical mass estimates of the Local Group (LG) are cosmology dependent. The TA is a method for calculating the mass of the LG, based on the assumption that most of the mass is bound in with the two largest galaxies: the Milky Way (MW) and Andromeda (M31). These are assumed to have originated from over-densities in approximately the same point in space in the early universe, which are then separated by the Hubble flow, and eventually turn around due to their gravitational interaction. In this model, the Milky Way and Andromeda have already passed turnaround and are heading for collision, on first approach, in the next 3-5 Gyr. Using the simple dynamics of the TA, and knowing the present separation of two galaxies, their radial relative velocity, and the age of the universe, we can calculate a unique estimate of the combined mass of the system. When one restricts oneself to the simple basis of the TA as an isolated, solely radially moving system, one finds that the mass estimate is dependent on the observed dynamics (distance and velocity) and the cosmology through the strength of dark energy (modifying the force law, as in \cite{Partridge}\cite{McLeod}) and the age of the universe (modifying the boundary conditions). Comparisons to simulations show that the TA can be an effective and unbiased estimate of the LG mass \cite{LW}\cite{McLeod}. 

The age of the universe is determined by fundamental cosmological parameters such as the matter density, the density of dark energy, and the Hubble parameter. Cosmology independent bounds on the age of the universe may be found from, for example, the age of globular clusters, however if one wishes to place tight constraints on the age of the universe (and, therefore, the mass of the LG), then one must use cosmological probes and models. The tension present between estimates of $H_0$ from the CMB, galaxy velocity flows, and LSS \cite{Planck}\cite{Hubble_Review} introduces uncertainty in the age of the universe, and thus also the TA mass estimate. Analyses based on $\Lambda$CDM simulations are generally unable to deal with these cosmological dependencies, since there is only a single realisation with fixed properties. It therefore does not take into account the uncertainty in the cosmology, and an estimate based on a different set of cosmological parameters or theory of dark energy would require retraining with an appropriately set up simulation. This is particularly stifling in the case of non-standard cosmologies such as scalar field models, where the properties of the field are poorly constrained and thus one would need to explore a wide parameter range, requiring an ensemble of detailed simulations.

We examine a variety of DE models and calculate the LG mass dependence on their parameters in the straightforward case of the TA (i.e. assuming purely radial motion). One can infer a cosmologically marginalised LG mass by using the posterior probability distributions from numerical sampling and cosmological data; the LG mass may in future be determined from joint observations of the LG dynamics, tidal streams, large scale structure, and even CMB constraints. Conversely, in principle at least, if one had a good independent estimate of the LG mass, one could use the LG as a `local laboratory' to constrain cosmology or gravity. These constraints are unlikely to be competitive with the CMB or large scale structure (LSS), but in the case of poorly constrained or highly scale dependent MG and DE models such local systems may be useful. Observations of systems such as the LG and other galaxy clusters could eventually lead to competitive DE constraints on small scales of $\mathcal{O}(\lesssim 1 \us \text{Mpc})$, which are not typically probed by cosmological analyses. This is relevant primarily for scale dependent theories such as scalar-tensor theories, but not for homogeneous theories such as $\Lambda$, perfect fluids, or spatially invariant scalar fields $\phi(t)$. There is a class of models for which such local study may be particularly useful: models which are designed to replace dark matter may be subjected to a consistency test within in the LG. The baryonic mass of each galaxy may be inferred from its rotation curve in the context of the given theory. This mass estimate ought to be consistent with the mass implied by the relative motion -- and thus gravitational interaction -- between the the two galaxies. If these mass estimates are significantly different, then a explanation must be found for the discrepancy. 

Estimates for the LG mass typically derive errors from observational uncertainties on the dynamical parameters (separation, relative velocity, etc.), assuming a particular cosmology. Is the uncertainty in the argument due to cosmological uncertainties comparable? A consistent cosmological approach to the LG mass may be taken by first calculating the background. This will give us the age of the universe $t_u$ from $a(t_u) = 0$ (defining the present as $t=0$). In addition we may obtain the full function $a(t)$, describing the entire expansion history of the universe at large. For simple models this may be calculated with the Friedmann equations for perfect fluid components. For more complicated models such as scalar-tensor theories one may require a cosmological Boltzmann code or a fixed background expansion. (We shall used a fixed $\Lambda$CDM background expansion for scalar-tensor models and MOND.)

When non-minimally coupled scalar fields are introduced, the argument no longer hold quite so simple a form. This is because the coupling to the scalar field is structure dependent -- the radius and density of the objects determines the scalar field profile around an object. As a result, we must abandon our assumption of halos are point particles in space, and the shape and density must be modelled. Typically in scalar tensor theories, the scalar field will behave in such a way as to produce little or no effect in areas of high density: this is known as `screening'. (It is frequently achieved by having the scalar field profile fall to zero at high density, or to have the effective mass of the scalar field become very large so that its interactions are unobservably short range.) Screening is a vital component to many theories in order to modify gravity at large scales, whilst avoiding violating tests which have been made on deviations from general relativity on solar system scales \cite{Brax}\cite{Mota}. Minimally coupled scalar-tensor models generically produce external solutions of the form $~ r^{-1} e^{-mr}$ under spherical symmetry in the perturbative regime (where variations in the scalar field are small compared to the background value and variations in the potential are small), with a theory and density dependent coupling term. In addition to this coupling, there is a dark energy like effect from the energy density of the scalar field; a constant $V_0$ in the potential will enter into the field equations as a cosmological constant. Scalar field theories are generally of interest to cosmologists as a model of dark energy, but fifth-force effects from scalar field theories can also be used to stabilise rotation curves without dark matter, as in \cite{Symmetron}. Although unified scalar field model of dark energy and dark matter is an attractive notion to some theorists, models have so far been tuned either to cosmological tests of dark energy or to local or galactic scale tests of gravity; to be successful, such models would have to be tested across all scales. Furthermore, if a scalar field is to serve as a replacement for dark matter, then the LG mass estimates should be consistent with the luminous mass of the galaxies. If it is not, then dynamics within our local universe could become a major hurdle for theories seeking to eliminate dark matter. Although a replacement for dark matter entirely is an extreme example, \cite{Symmetron} nevertheless demonstrates that the presence of a minimally coupled scalar field can alter estimates of the dark matter distribution within a galaxy when scalar field fifth-forces are taken into account. It is therefore useful to have a framework in place to analyse the dynamical impact of scalar field theories, whatever the motivation behind their introduction. 

The outline of the paper is as follows: section 2 summarises the properties of the LG as they pertain to this work. Sections 3-5 look at the impact of different dark energy theories and cosmological parameters on the LG mass estimate, in which section 3 explores the TA with $\Lambda$ and a $\Lambda$CDM universe, section 4 explores a perfect fluid model of dark energy and a $w$CDM universe, and section 5 turns to a minimally coupled scalar field model with a time dependent but spatially invariant scalar field. Sections 6 and 7 look at using the TA to estimate the mass of the LG in models which were constructed to explain rotation curves of galaxies without dark matter. Section 6 looks at coupled scalar field solutions and a dark matter free symmetron model, and section 7 looks at a another alternative to dark matter, MOND. There is a discussion of the results in section 8.  

\section{Properties of the Local Group}

The TA requires the relative radial separation and velocity of MW and M31, which we take to be \cite{vdMarel}
\begin{align}
r &= 0.77 \pm 0.04 \us \text{Mpc}, \\
v_r &= -109.4 \pm  4.4 \us \text{km}\us{s}^{-1},
\end{align}
$r$ is the separation and $v_r$ is the relative velocity along the separation vector, and we have also quoted $1\sigma$ errors. We note that a new analysis, based on data from Gaia and the Hubble Space Telescope, of the LG \cite{VdM19} produces a transverse velocity estimate of $v_ t = 57^{+35}_{ -31} \us \text{km}\us{s}^{-1}$, which is larger than previously estimated and challenges the well established assumption of a radial orbit; the implications of this new value are studied elsewhere \cite{Benisty}. We will deal predominantly in units of Mpc, Gyr, and masses will be presented as multiples of $10^{12} M_\odot$. We will assume that $v_t$ is consistent with zero for this paper. 

We should also note that, although the TA simplifies the structure to two points masses, the mass which it is modelling will include any extended structures which are gravitationally bound to, and move with, the MW and M31. This means the masses may be higher than those from other estimates of the galaxies individually. 

In order to calculate $M_{LG}$ in scalar-tensor theories, which depend on the structure of the objects and not just their mass, we need to select some additional properties for our system relating to the radius and mass of the individual galaxies. In order to calculate the gravitational interactions in scalar field theories, we need not only the total mass, but also the individual masses, densities, and shape of the halos in order to determine their individual scalar field profiles. We shall make of the approximation of uniform spheres, and thus this information reduces to the individual masses and radii of two halos. This is the simplest extension to the TA's assumption of a point mass, and simplifies the scalar field solution. This should provide a first approximation for the effects of the scalar field on the dynamics, particularly given that for the majority of the orbits the separation is much greater than the size of the structures, but more thorough investigations of this model require detailed numerical simulations of the structures of the galaxies and their satellites. 

As fiducial values we take $M_{M31} = 2 M_{MW}$ for the masses of the two galaxies, as suggested by studies such as \cite{Phelps}; the LG mass is the sum $M_{LG} = M_{M31} + M_{MW}$. The radii of the galaxies will be $r_{M31} = 33\us\text{kpc}$, and $r_{MW} = 30\us\text{kpc}$ for their visible extent. These values are however uncertain, particularly the mass ratio of the two galaxies, and will be varied in the section on scalar tensor theories. The radii will be varied between 20 and 40 kpc, which contains most estimates of the radii of the galaxies. For dark matter free theories, the luminous mass of the MW and of M31 should be $\mathcal{O}(10^{11}M_\odot)$, with the original TA paper \cite{KW} suggesting masses of $10^{11} M_\odot$ and $4\times 10^{11} M_\odot$ for MW and M31 respectively, and a more recent paper exploring a dark matter-less MW arriving at $2.7\times 10^{11} M_\odot$ \cite{MW_luminous}. We will therefore expect dark matter free theories to produce a mass in approximately this range in order to be consistent with observations. 

\begin{table*}[t]
\renewcommand{\arraystretch}{2}
\centering
\caption{A summary of the models used in this paper, their free parameters, and the section in which they are found. $\ddot{r}_N$  denotes the standard Newtonian acceleration ($\frac{-GM}{r^2}$). }
\label{Model_Table}
\begin{tabular}{lllll}
\hline
Model & Acceleration Equation & Free Parameters & Section \\ \hline
$\Lambda$CDM & $\ddot{r} = \ddot{r}_N + \frac{\Lambda c^2}{3} r$ & $\Omega_\Lambda$, $H_0$ & \ref{TA_LCDM} \\ \hline
$w$CDM & $\ddot{r} = \ddot{r}_N - \frac{1}{2}H_0^2{\Omega_f} (1+3w) R(t)^{-3(1+w)}r$ & $\Omega_f$, $w$, $H_0$ & \ref{TA_wCDM} \\ \hline
Quintessence & $ \ddot{r} = \ddot{r}_N + \frac{4\pi G}{3}[\dot{\phi}^2 + 2V(\phi)]r$ & $\Omega_\phi$, $\dot\phi_0$, $\phi_*$, $H_0$ & \ref{TA_Quint} \\ \hline
Symmetron & $ \ddot{r} = \ddot{r}_N + 4\pi \frac{C_AC_B}{M} \frac{e^{-mr}}{mr} \left( 1 + \frac{1}{mr} \right)$ & $\mu$, $\lambda$, $S$ & \ref{TA_Symm} \\ \hline
MOND & $\ddot{r} = - \left[ \frac{\ddot{r}^2_N + \sqrt{ \ddot{r}^4_N + 4\ddot{r}_N^2a_0^2}}{2} \right]^{\frac{1}{2}}$ & $a_0$, $t_u$ & \ref{TA_MOND} \\ \hline
\end{tabular}
\end{table*}

\section{The TA in $\Lambda$CDM}

The original Timing Argument (derived in \cite{KW}) augmented with a cosmological constant (in accordance with the weak field limit of GR+$\Lambda$, see \cite{Partridge}, \cite{BaT}), hereafter referred to as $TA_\Lambda$, is highly cosmological in nature. It assumed that the MW and M31 form in close proximity in the early universe and move with the expansion of the universe, later turning around under the influence of gravity to reach their present configuration. The mass is determined from the age of the universe, since the spatial coincidence of the galaxies ($r = 0$) occurs at $t_* = t_0-t_u$, where $t_0$ is the present time and $t_u$ is the age of the universe. The dynamics are governed by a Newtonian equation with an additional $\Lambda$ term, 
\begin{equation}\label{TA_Lambda}
\ddot r = - \frac{GM}{r^2} + \frac{\Lambda c^2}{3} r 
\end{equation}
which may be obtained by taking the weak field limit of the Einstein field equations with a cosmological constant. This is the acceleration equation which we must use to calculate the LG mass if dark energy (in its simplest form) is to be taken properly into account. 

\subsection{TA$_{\Lambda}$ as an Isolated Patch}

We can also understand this equation by noting that, since the matter density $\rho$ considered in deriving this contained only the point particles under consideration, this implicitly assumed that, locally, we are in a patch containing only the galaxy pair (comprising all of the mass) and a background of dark energy. We will show that the equation of motion for two particles moving under gravity, in a dark energy filled background evolving according to the Friedmann equations, is identical to the form in equation \ref{TA_Lambda}. 

To begin with, consider the equations in the comoving coordinate $x$ for a particle on a background where the physical distance $r = a(t)x$ (see \cite{Peebles})
\begin{equation}\label{comoving}
\ddot x + 2 H_* \dot x + \frac{1}{a^2}\frac{\partial \Delta\Phi}{\partial x} = 0,
\end{equation}
where $\Delta \Phi$ represents the contribution to the gravitational potential from the particles, not including the background (the effects of which are absorbed into the scale factor $a$). We have used $H_*$ for the Hubble parameter as a reminder that this is not $H$ for a standard $\Lambda$CDM universe, but applies to an empty background with only $\Lambda$. We will likewise use $a_*$ for the scale factor. 
We can find the gravitational force using the comoving Poisson equation for a spherical density distribution and a background density which is stationary w.r.t. to the comoving coordinates:
\begin{equation}
x^{-2} \partial_x (x^2 \partial_x \Delta\Phi) = 4\pi G (\rho - \rho_b)
\end{equation}
\begin{align}
 \partial_x \Delta\Phi &= x^{-2} 4\pi G \int (\rho - \rho_b) x^2 \mo{dx} \\
					&= x^{-2} 4\pi G a_*^{-1} \int (\rho - \rho_b) r^2 dr \\
					& = \frac{GM}{a_*x^2}
\end{align}
which gives the equation of motion in comoving coordinates for the overdensity 
\begin{equation}\label{comoving}
\ddot x + 2 H_* \dot x + \frac{GM}{a_*^3x^2} = 0
\end{equation}
Since the matter is bound up in our particles, locally the background is made up on only dark energy. From the cosmological field equations, in a universe with only a cosmological constant
\begin{equation}
H_*^2 = H_{*0}^2 \Omega_{\Lambda *}= \frac{\Lambda c^2}{3} = \text{const}
\end{equation}
Using $\ddot r = a_* \left[H_*^2 x + 2H_*\dot x + \ddot x\right]$, and substituting in $\ddot{x}$ from equation \ref{comoving}, we can convert the comoving equation to our usual physical equation in $r$:
\begin{equation}
\begin{split}
\ddot{r} &=  -\frac{GM}{r^2} + H_*^2r \\
	 	& = -\frac{GM}{r^2} + \frac{\Lambda c^2}{3}r
\end{split}
\end{equation}
which is the same as equation \ref{TA_Lambda}, even though there is no $\Lambda$ term in the comoving acceleration equation. The effects of dark energy, which in $\Lambda$CDM are constant, are absorbed entirely into the scale factor $a(t)$; this is the common way to express the equations in n-body simulations and similar applications. The effect of the cosmological constant can be seen from comparisons of TA and TA$_\Lambda$ with the simulations, as in \cite{McLeod}.

It is important to note that this is not the evolution of the universe as a whole, which on large scales is comprised of a smooth distribution of matter and radiation, in addition to dark energy; the scale factor $a(t)$ here is very different from the scale factor of a full $\Lambda$CDM universe. This patch of universe evolves quite independently of the background expansion which governs the observable universe as a whole (it is `decoupled' from the Hubble flow). The `timing' aspect thus relates to the fact that the evolution of this patch of the universe must be consistent with the initial conditions of the universe as a whole i.e. the origin at $t = t_0-t_u$. Although the TA considers the system in isolation from the rest of the universe, the value of $\Lambda$ must be the same since it is a fundamental constant.  

\subsection{The LG Mass in $\Lambda$CDM}\label{TA_LCDM}

In $\Lambda$CDM, the only two cosmological parameters which affect the TA$_\Lambda$ are $h$ and $\Omega_\Lambda$ (assuming that a flat universe and negligible contributions from relativistic species, and therefore that $\Omega_\text{m} = 1 - \Omega_\Lambda$). These values are used to calculate the age of the universe $t_u$, as well as to calculate the value of $\Lambda$ required for the dynamics of the galaxy pair. Note therefore that the cosmological parameters $h$ and $\Omega_\Lambda$ here are for the universe as a whole rather than the galaxies' isolated system, the dynamics of which are calculated using equation \ref{TA_Lambda}.

\begin{figure} 
  \caption{Mass estimate contours for the LG using $\Lambda$CDM. The mass is given in units of $10^{12}M_\odot$. The dependence of the mass on the cosmological parameters comes from the impact on $t_u$ and the $\Lambda$ modification to the acceleration equation. The black dot represents the mass estimate using $\Omega_\Lambda = 0.7$ and $h = 0.67$, which is a typical mass estimate for a Planck-like cosmology.}
   \label{LG_OL-h}
\includegraphics[width=8cm,keepaspectratio=true]{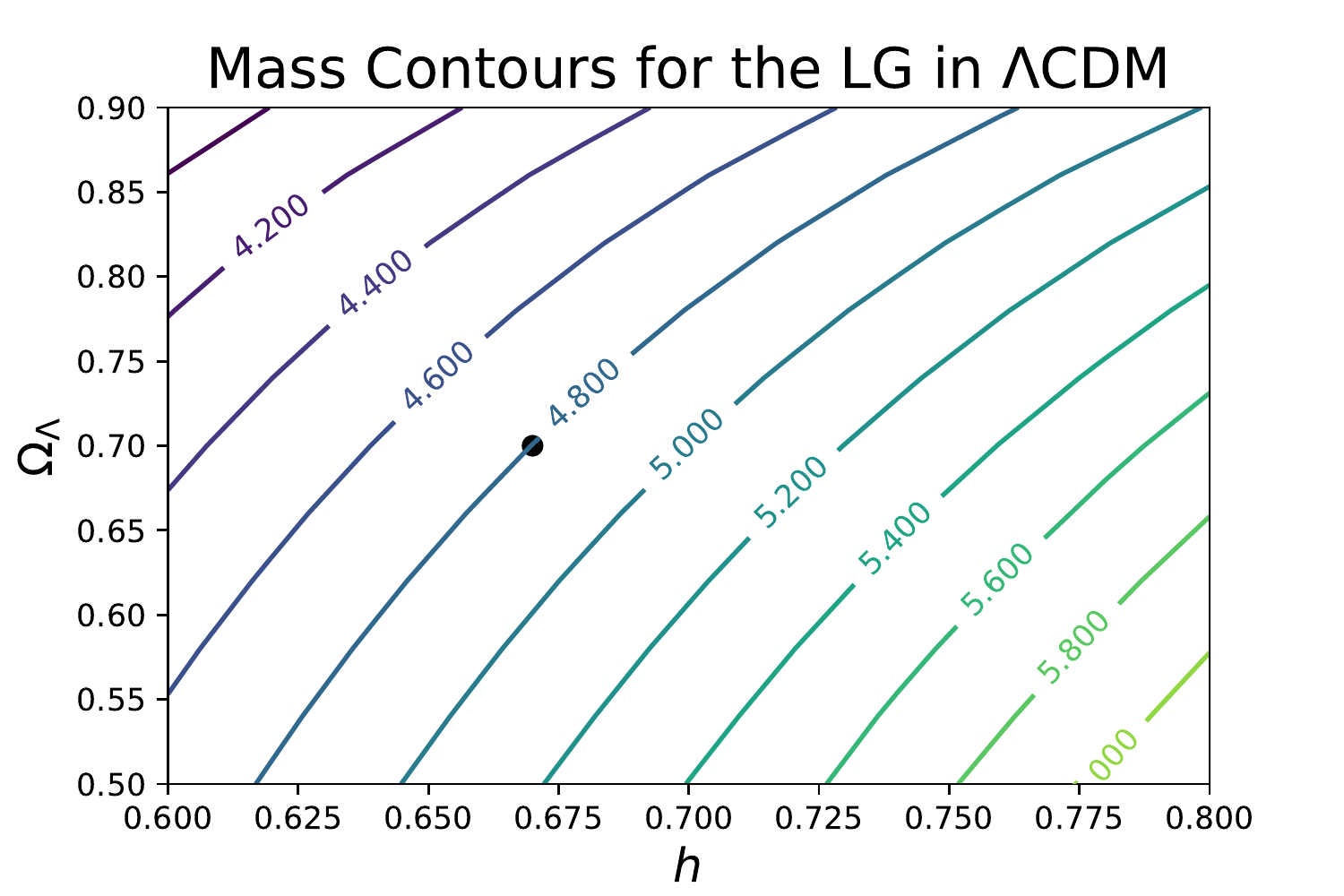}
\end{figure}

Fig \ref{LG_OL-h} shows the LG mass estimate contours for $\Lambda$CDM varying the present values of $\Omega_\Lambda$ and $h$. The effect of $\Omega_\Lambda$ appear different than in earlier work. In \cite{Partridge} and \cite{McLeod} one sees an increase in mass when $\Lambda$ is introduced into the TA (keeping $t_u$ constant) since this creates a repulsive effect in the acceleration equation \ref{TA_Lambda}; this repulsion must be overcome by an increase in mass relative to the standard Newtonian case in order to recover the same observed dynamics. However here assume $\Lambda$ as part of the dynamics, and we vary $\Omega_\Lambda$, which affects both the strength of the $\Lambda$ effect in equation \ref{TA_Lambda} and the age of the universe $t_u$. Since an increased $\Omega_\Lambda$ leads to an older universe, the interaction must be weaker in order to avoid collapsing too soon: this leads to the mass estimate falling rather than rising. This effect is dominant over the repulsive effect of $\Lambda$. This demonstrates the importance of using the cosmological parameters in a consistent way in order to see the genuine effect on the dynamics. The effect of $h$ on the mass estimate is more obvious. The age of the universe varies as $\sim H_0^{-1}$, and thus as our mass estimate should increase as $h$ increases in order to ensure a fast enough collapse. $\Omega_\Lambda$ and $h$ are also not independent, since 
\begin{equation}
\Omega_\Lambda = \frac{\Lambda c^2}{3H^2}
\end{equation}
and thus it is $H^2\Omega_\Lambda$ which enters into the acceleration equation. Nevertheless, $\Omega_\Lambda$ is much more useful for cosmology than $\Lambda$ itself, and it is more useful to look at the the impact of the parameters which are widely used. We can see from Fig \ref{LG_OL-h} that varying $h$ over the range of different estimates made by various cosmological probes could yield significant changes to the LG mass estimate. Taking $h$ in the range $[0.632,0.764]$ (to take the extremes of \cite{Hubble_Review}, the lowest value being the lower bound for a study of Megamasers \cite{Megamaser}, and the higher value being the upper bound for a Cepheid study \cite{Cepheid}), and a fixed $\Omega_\Lambda = 0.7$, we find the mass falls in the range $[4.55 \times 10^{12}M_\odot, 5.43\times 10^{12}M_\odot]$. 

\section{Timing Argument for Perfect Fluids}\label{TA_wCDM}

To cast the TA for different theories of gravity, we need to look at the weak field limit of general relativity including the relevant dark energy model. Perfect fluid models with a constant equation of state have one extra free parameter, which is their equation of state $w = \frac{P_f}{\rho_f}$, where $w_\Lambda = -1$. The density of DE ($\rho_f$) and the equation of state will both affect the acceleration equation. The equation of state enters into both the energy-momentum tensor (which therefore affects the gravitational strength at a given time) and also the variation in density of dark energy (which is constant for $\Lambda$). 
 
The weak field quasi-static (non relativistic) limit of general relativity gives us 
\begin{equation}\label{GR_fluid}
\nabla^2 \Phi = 4\pi G (\rho_m + (1+3w)\rho_f)
\end{equation}
Since $\nabla^2$ is a linear operator, we can split the potential into two parts $\Phi_m$ and $\Phi_\text{DE}$, which correspond to the matter and dark energy potentials; these then lead to separate contributions to the acceleration equation $\ddot{r} = \ddot{r}_m + \ddot{r}_\text{DE}$. The matter potential leads to the familiar Newtonian expression for a spherical density distribution. The contribution to the acceleration equation can therefore be determined by integrating only the fluid part of the equation:
\begin{equation}
\nabla^2 \Phi_\text{DE} = 4\pi G (1+3w)\rho_f
\end{equation}
We will assume that $w$ is constant, and that $\rho_f$ is not spatially varying (i.e. dark energy is non-clustering). 
The contribution to the acceleration is therefore:
\begin{equation}
\ddot{r}_\text{DE} = -\nabla \Phi_\text{DE} = - \frac{4 \pi G (1+3w)}{r^2} \int_0^r \rho_f(r^\prime) {r^{\prime}}^2 \mathop{dr^\prime}
\end{equation}
which depends on the equation of state and the integrated energy density of the dark energy fluid. The case of a cosmological constant is easily recovered when $w = -1$ and $\rho_f = \rho_\Lambda = \frac{\Lambda c^2}{8\pi G}$. Integrating, we have:
\begin{equation}
\ddot{r}_\text{DE} = -\frac{4\pi G}{3} (1+3w) \rho_f r
\end{equation}
Since our dark energy is not clustered, the density is determined entirely by the background of the universe at large. 
\begin{equation}
\rho_f = \rho_{f,0}\us a^{-3(1+w)}
\end{equation}
The acceleration equation is therefore:
\begin{equation}\label{w_acceleration}
\ddot{r} = -\frac{GM}{r^2} - \frac{1}{2}H_0^2{\Omega_f} (1+3w) a(t)^{-3(1+w)}r
\end{equation}

In order to consistently use dark energy models in the TA, we must also know how it affects the age of the universe and, in the case of a perfect fluid, its expansion history $a(t)$. Both can be readily calculated from the Friedmann equations using cosmological parameters $H_0$ and $\Omega_\text{DE,0}$ (assuming $\Omega_\text{m,0} = 1 - \Omega_\text{DE,0}$) to set the conditions at the present time, and then integrating back until $a(t_0 - t_u) = 0$. 
\begin{equation}
H^2 = H_0^2 \us\sum_i \left[ \Omega_{i,0} \us a^{-3(1+w_i)} \right]
\end{equation}
There is an analytic expression for a(t) for the case of a two component (matter and dark energy) model with constant $w$:
\begin{equation}
H_0 t = \frac{2 \ln\left( \sqrt{\Omega_f (\Omega_m + \Omega_f a^3)} + \Omega_f a^\frac{3}{2} \right)}{3 \sqrt{\Omega_f}} - \frac{2 \ln(\sqrt{\Omega_m\Omega_f})}{3 \sqrt{\Omega_f}}
\end{equation}
The age of the universe is the difference in time between $a = 1$ and $a = 0$. (Alternatively, it can be readily obtained by numerical integration.) Again we shall neglect contributions from relativistic species and other cosmological parameters, since these have a small effect on $t_u$, and the effect upon the mass from this will be negligible compared to other sources of uncertainty. 

The mass is then numerically calculated from $(r,v_r,t_u)$, using this background cosmology to inform the acceleration equation \ref{w_acceleration}. This is done by calculating $t$ (the time since $r = 0$) for different values of $M$ (given $r$, $v_r$, $w$, and $\rho_f$) using numerical integration, and using a root finding algorithm to find the $M$ such that $t = t_u$. 

\begin{figure} 
  \caption{Mass estimate contours for the LG using a spatially homogeneous perfect fluids with constant $w$. The black dot corresponds to $\Lambda$CDM with $\Omega_\Lambda = 0.7$ ($\Omega_f = 0.7$, $w = -1$) and $h = 0.67$. }
   \label{LG_wH}
\includegraphics[width=8cm,keepaspectratio=true]{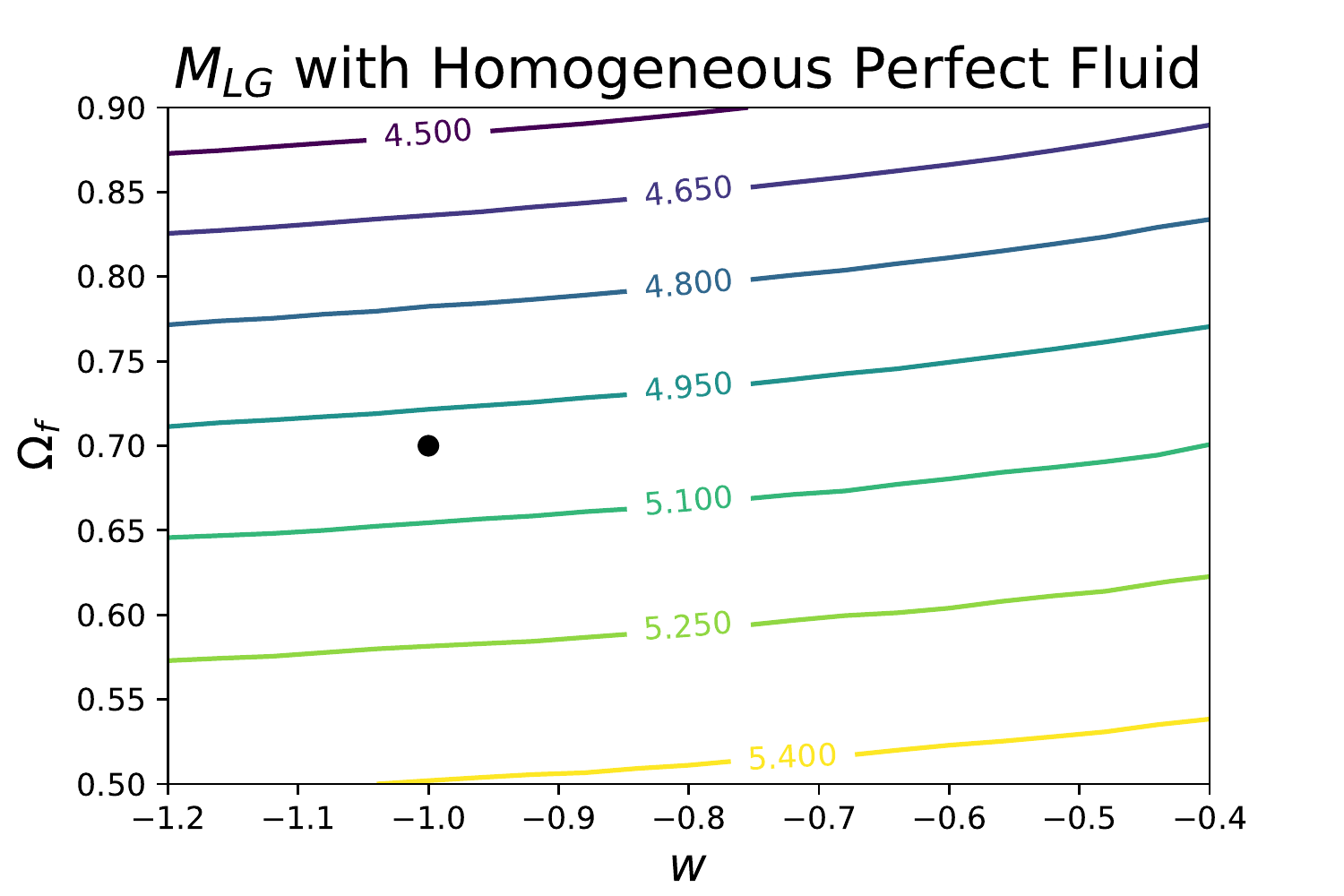}
\includegraphics[width=8cm,keepaspectratio=true]{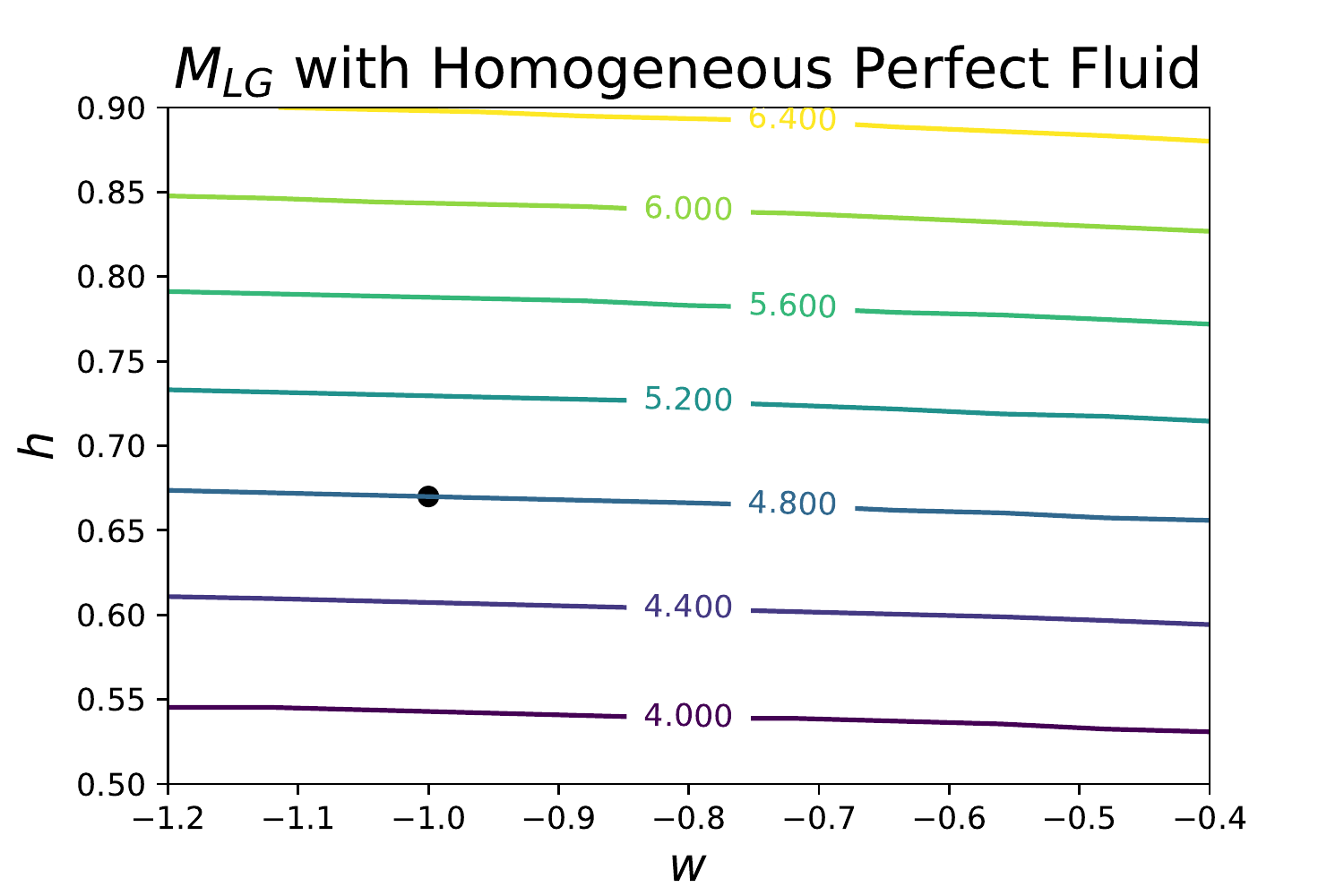}
\end{figure}


Figs \ref{LG_wH} shows the LG mass estimate as a function of $\Omega_{f}$, $w$, and $h$ for a spatially homogeneous perfect fluid with constant equation of state. The LG mass is not very sensitive to the equation of state at all, even when deviations from $\Lambda$CDM are extreme. Like the case with $\Lambda$, there are two opposing effects in play here. Decreasing $w$ decreasing the age of the universe, but also diminishes the dark energy expansion effect in the acceleration equation. These effects are largely canceling each other out in the mass calculation. 

\section{Minimally Coupled Scalar Field}\label{TA_Quint}

A scalar field which does not directly couple to gravity or to matter can produce a dark energy effect. If it has only time dependence then it can be easily inserted into the gravitational equations using the energy-momentum tensor of the scalar field. The scalar field evolves according to a potential function $V(\phi)$; this must be tuned to produce the effect that we want. Minimally coupled scalar fields are typically invoked in inflation theories, although they also form the basis of quintessence theories of dark energy. Quintessence was introduced in \cite{Quintessence_Intro}, and a review can be found here \cite{Quintessence_Review}. In our case we will make use of straightforward properties of the scalar field which may be found in textbooks such as \cite{Hobson}. Attempts to constraint the quintessence potential have been made in \cite{Quintessence_Potential}, although the uncertainties remain large, and higher than linear or quadratic terms in the potential remain unconstrained. 


Since we are assuming a spatially homogeneous scalar field across the entire universe, the scalar field value (like the density of the perfect fluid in the previous section) must be calculated taking the evolution of the universe into account. Using the Klein-Gordon equation:
\begin{equation}
\Box \phi = g^{\mu\nu} \nabla_\mu \nabla_\nu \phi = \partial_\phi V(\phi) 
\end{equation}
and using the FLRW metric and the fact that $\phi$ is a function of $t$ only, one obtains:
\begin{equation}
\ddot \phi + 3H\dot\phi - \partial_\phi V(\phi) = 0
\end{equation}
In order to calculate the evolution of the scalar field then, we need to have $V(\phi)$, $\phi_0$, and $\dot \phi_0$.

\begin{figure} 
  \caption{Mass estimate contours for the LG with a spatially homogeneous scalar field with an exponential potential $V(\phi) = V_0 e^{-\frac{\phi}{\phi_*}}$. The dashed line represents $\dot\phi = 0.0$, which approached $\Lambda$CDM in the limit $\phi_* \rightarrow \infty$. Scalar field quantities are presented in Planck units.}
   \label{LG_MC}
\includegraphics[width=8cm,keepaspectratio=true]{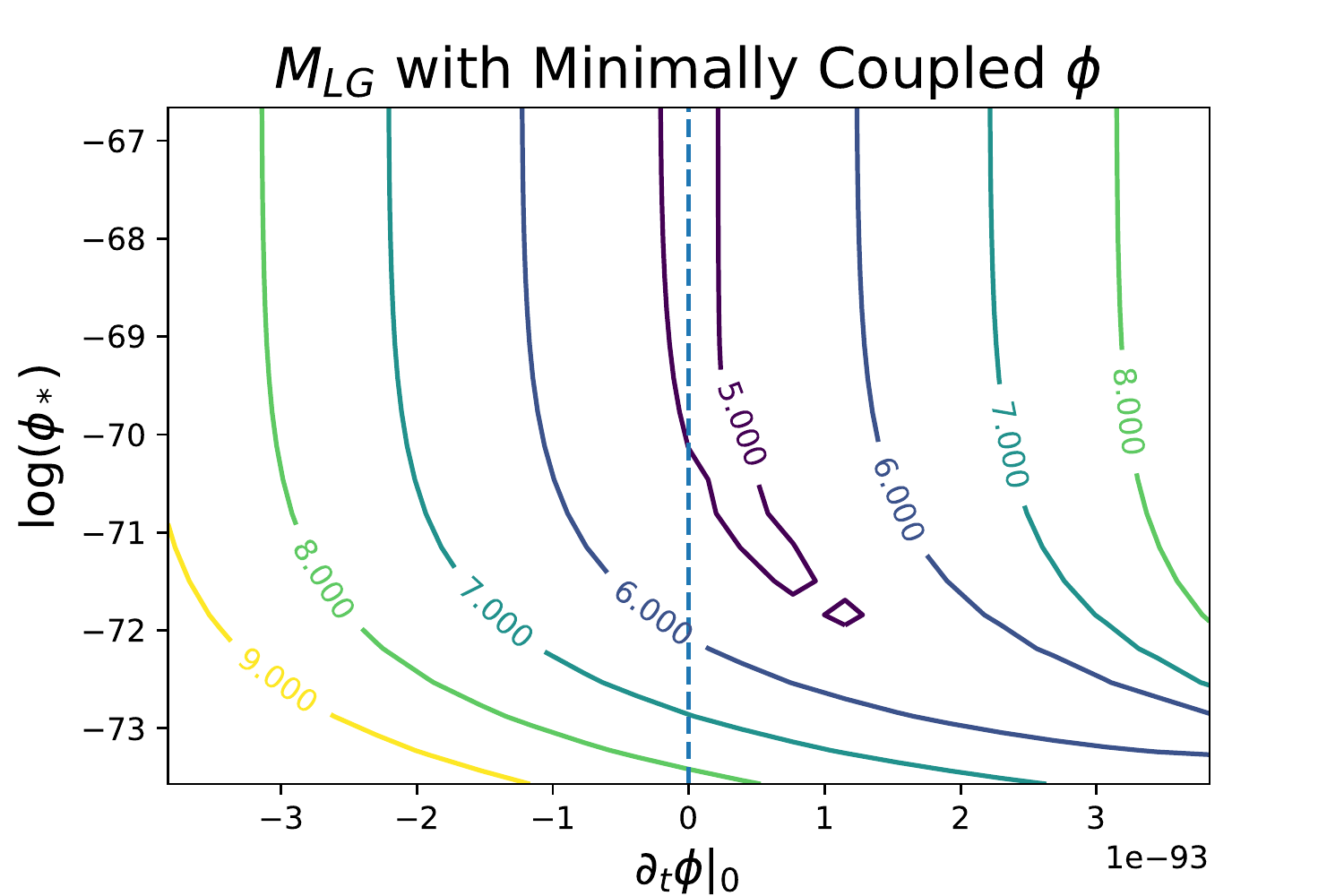}
\end{figure}

The acceleration equation for a universe with a scalar field can be found using the expression for the scalar field energy momentum tensor, which is itself obtained by varying the standard action for a minimally coupled field with respect to the metric:
\begin{equation}
T^\phi_{\mu \nu} = \partial_\mu \phi \partial_\nu \phi - g_{\mu \nu} \left( \frac{1}{2} \partial_\sigma \phi \partial^\sigma \phi - V(\phi) \right)
\end{equation}
Inserting this into the Einstein field equations gives the acceleration equation. Using the Newtonian limit (setting $c = 1$)
\begin{equation}\label{Poisson_scalar_min}
\nabla^2 \Phi = 4\pi G\left[ \rho - 2\dot\phi^2 - 2V(\phi) \right]
\end{equation}
It is easy to show that this is equivalent to inserting $\rho_\phi - 3 p_\phi$ into equation \ref{GR_fluid} i.e. that the field in this limit acts as a perfect fluid with variable $w_\phi$. The limiting case of a cosmological constant found when $\dot\phi = 0$ and therefore:
\begin{equation}
8\pi GV(\phi) = \Lambda
\end{equation}
Integrating equation \ref{Poisson_scalar_min} and bearing in mind that the scalar field has no spatial dependence, one obtains
\begin{equation}
\ddot r = -\frac{GM}{r^2} + \frac{8\pi G}{3}\left[ \dot\phi^2  + V(\phi) \right] r
\end{equation}
We use an exponential form for the quintessence potential, choosing:
\begin{equation}
V(\phi) = V_0 e^{-\frac{\phi}{\phi_*}}
\end{equation}
We may set $\phi_0 = 0$ without loss of generality, as the effect of starting $\phi$ at different points can be absorbed into $V_0$. $\phi_*$ sets the gradient of the potential, and $\dot\phi$ sets its initial trajectory. We calculate $V_0$ by relating it to $\Omega_\phi$:
\begin{equation}
\Omega_\phi = \frac{8\pi G \rho_\phi}{3H^2} \implies V_0 = \frac{3H^2\Omega_\phi}{8\pi G} - \dot\phi_0^2
\end{equation}
With these initial conditions, the model is fully governed by $\Omega_\phi$, $\phi_*$, and $\dot\phi_0$. Fig \ref{LG_MC} shows the variation in the mass estimate with $\dot\phi_0$ and $\phi_*$. We hold $\Omega_\phi = 0.7$ in order to reproduce the $\Omega_\Lambda = 0.7$ $\Lambda$CDM limit, and keep $h_0 = 0.67$. We can see that the LG mass is insensitive to $\phi_*$ when the parameter is large enough. This is because the potential is flattened, and its contribution becomes roughly constant for all $\phi_* \gg \phi$. In this case the change to the mass is also almost symmetric in $\dot\phi$. The field velocity only enters into the acceleration equation as $\dot\phi^2$, so if the potential is extremely flat and thus making little impact on the field velocity then positive and negative velocities will be almost indistinguishable. When $\phi_*$ is small enough, then the potential is rapidly varying and it becomes the dominant part of the dynamics, with the initial scalar field velocity $\dot\phi_0$ quickly becoming overwhelmed by the roll down the potential. 

\section{Coupled Scalar Fields}\label{TA_Symm}

The scalar field need not act independently of the matter. Coupled scalar fields -- scalar-tensor theories -- interact with the matter or the spacetime curvature directly to produce spatially inhomogeneous fields. A thorough introduction to scalar-tensor theories can be found in \cite{Peter}, and a broad but less detailed review in \cite{Brax}. 

The Newtonian limit of such a scalar field theory is determined by the matter coupling $\Omega(\varphi)$ and the evolution of the scalar field. For a given theory the scalar field is determined by a Klein-Gordon equation with an effective potential which, in the Einstein frame, depends on the matter density. The masses and radii of MW and M31 are likely different, and therefore the profile of the scalar field around each of them may be different. Therefore we cannot combine their mass in the acceleration equations quite so simplistically, and the equation will involve both masses individually. 

Scalar fields may be used to model either dark energy or dark matter (as in the symmetron field in \cite{Symmetron}) which means that some scalar-tensor theories may be comparable to MOND. The energy density of the scalar field can act in the same way as a cosmological constant and produce a universal repulsion; this is true even for fields which produce an increased gravitational potential which can mimic dark matter. It would certainly be interesting if the dark aspects of cosmology could both be illuminated by the additional of a single scalar field! In general however, the potential of a scalar field theory must often be augmented with an additional constant to mimic $\Lambda$.

\subsection{The Klein Gordon Equation for Scalar Fields}

Perhaps the first thing we need to understand is the field equation for the scalar itself. This is often referred to as the `Klein-Gordon equation' of the scalar field, by analogy with the famous equation for free fields. The field will be subject to some effective potential which, in the case of coupled scalar fields, will be a function of the scalar field and either the curvature (Jordan frame) or the matter density (Einstein frame). 
Let us assume we have a scalar field $\phi$, which is subject to a scalar field equation:
\begin{equation}
\Box^2 \phi + \partial_\phi V_\text{eff}(\phi) = 0
\end{equation}
Provided that the effective potential $V_\text{eff}$ has a minimum at some $\phi_0$, then we can say 
\begin{equation}
\partial_\phi V_\text{eff}(\phi_0) = 0
\end{equation}
We can expand the scalar field around the minimum such that $\phi = \phi_0 + \varphi$. The linearised equation then reads:
\begin{equation}
\Box^2 \varphi  + \partial^2_\phi V_\text{eff}(\phi_0) \varphi = 0 
\end{equation}
We can then identify $\partial^2_\phi V_\text{eff}(\phi_0)$ with an effective mass $m^2_{\phi_0}$, and rewrite the equation as
\begin{equation}\label{KGeqn}
\left( \Box^2 + m_{\phi_0}^2 \right) \varphi = 0 
\end{equation}
which makes the connection to the Klein-Gordon equation. We can already see a key feature of coupled theories: the scalar field mass become density (or curvature) dependent. This is what allows the scalar field to `hide' in high density environments, evading laboratory or solar system detection. 

\subsection{Generic Spherically Symmetric Solutions for Uniform Spheres}

Spherically symmetric solutions for the scalar field equation have been discussed in \cite{Brax}\cite{Mota}\cite{Chameleon}. We develop a solution along similar lines to \cite{Chameleon} in the case where the equations can be linearised everywhere. Equation \ref{KGeqn} for the scalar field perturbation will in general have complex wave like solutions, although in the quasi-static limit there are growing and decaying modes. We shall find solutions of the form
\begin{equation}\label{trial}
\varphi = C x^a e^{bx}
\end{equation}
for real $C$, $a$, and complex $b$ (since $\varphi$ is a real scalar field). In the quasi-static case where we ignore time derivatives, $\Box^2 \rightarrow - \nabla^2$, and we may expand equation \ref{KGeqn} for a spherically symmetric case as:
\begin{equation}
\varphi^{\prime \prime} + \frac{2}{r} \varphi^\prime - m^2_{\phi_0} \varphi = 0 
\end{equation}
where prime indicates derivatives with respect to $r$. If we change coordinates to $x = m_{\phi_0} r$, then the equation can be written:
\begin{equation}\label{spherical}
\varphi^{\prime \prime} + \frac{2}{x} \varphi^\prime - \varphi = 0
\end{equation}
where prime now indicates derivatives with respect to the dimensionless scaled coordinate $x$. Using our trial solution:
\begin{align}
\varphi &= Cx^a e^{bx} \\
\varphi^\prime &= \left[ \frac{a}{x} + b \right] \varphi \\
\varphi^{\prime \prime} &= \left[ \frac{a^2 - a}{x^2} + \frac{2ab}{x} + b^2 \right] \varphi
\end{align}
Inserting these expressions into equation \ref{spherical} and collecting terms in factors of $x$ yields the consistency equations:
\begin{align}
b^2 -1 &= 0 \\
a(a+1) &= 0 \\
2ab + 2b &= 0
\end{align}
which can be solved by $a = -1$ and $b = \pm 1$. We are then generically led to solutions of the form:
\begin{equation}
\phi = \phi_0 + \varphi = \phi_0 + \frac{a}{m_{\phi_0} r} \left[ A e^{-m_{\phi_0} r} + B e^{m_{\phi_0} r} \right]
\end{equation}
This solution can be applied around the minimum of a generic potential form. It is the boundary conditions which lead us to the different internal and external solutions. At the centre we must have, by spherical symmetry, $\varphi^\prime = 0$. We also insist that the scalar field should tend to the free space minimum of the potential $\phi_{bg}$ in the limit of large $r$, and thus our functional forms are determined by:
\begin{equation}
\varphi^\prime(r=0) = 0
\end{equation}
\begin{equation}
\lim_{r\rightarrow \infty} \varphi(r) = \lim_{r \rightarrow \infty} \varphi^\prime(r) = 0
\end{equation}
This leads to solutions of the form
\begin{equation}\label{spherical_field_solutions}
\phi = \begin{cases}
		\phi_c + D \frac{\sinh(m_c r)}{m_c r} & \text{\emph{internal}} \\
		\phi_{bg} - C \frac{\exp(-m_{bg} r)}{m_{bg} r} & \text{\emph{external}}
	\end{cases}
\end{equation}
Where $\phi_c$ is the scalar field value which minimises the internal potential $\rho \neq 0$, and $\phi_{bg}$ is the field value which minimises the potential outside. 
The crucial constant $C$ -- which determines the amplitude of the scalar field perturbation far from the object and therefore also the modification to gravitational attraction -- may be determined by matching to the internal solution to the external solution at the boundary of the object. Fixing both $\phi$ and $\phi^\prime$ fixes both $D$ and $C$ (although $D$ holds no real significance on its own). The obtained coupling strength, $C$, will depend on $M_\text{LG}$ since the internal solution depends on the structure of the object including its density and radius. 

Considering a uniform sphere with radius $R$, we can join the solutions at the boundary of the sphere. This leads to an amplitude of 
\begin{equation}\label{coupling}
C = (\phi_{bg} - \phi_c) e^{m_{bg}R} \left[ \left(1 + \frac{1}{m_{bg}R}\right) \frac{1}{m_cR \coth(m_cR) -1} + \frac{1}{m_{bg}R} \right]^{-1}
\end{equation}
In the case where $m_{bg}R, m_cR \ll 1$ we may Taylor expand the above expression to arrive the simpler (and more intuitive) expression:
\begin{equation}\label{couping_simplified}
C \approx \frac{\Delta \phi}{3} x_{bg} x_c^2
\end{equation}
where $x_{bg} = m_{bg}R$, $x_c = m_cR$, and $\Delta \phi = \phi_{bg} - \phi_c$. 
Although this is only valid in the perturbative regime for a highly idealised object, it can help us understand how the coupling to objects depends on the internal structure through changes to the minimum ground state and effective mass of the scalar field inside the object, which in turn depends on its density and radius. 

\subsection{Solutions for Strongly Perturbed Scalar Fields in Uniform and Non-Uniform Spheres}

Should we wish to relax our assumptions and calculate solutions for more complex object or a more strongly perturbing regime, we must resort to numerical analyses in the general case. Numerical methods are used in \cite{Symmetron} to calculate the field inside a non-spherical galaxy, and \cite{Chameleon} considers strongly perturbing solutions. In order to numerically integrate the solution, one must then set the initial values of the scalar field at the centre of the object to be: 
\begin{align}
\phi(r = 0) &= \phi_* \\
\phi^\prime(r = 0) &= 0
\end{align}
Different values of $\phi_*$ will lead to radically different behaviours, and we seek the numerical solution which obeys the asymptotic conditions outlined above in the analytic solution. We know that at large $r$ we must have the exponentially decaying solution if we have a constant background density (which we may take to be zero) since we will eventually be close enough to $\phi_{bg}$ for the perturbative solution to be valid. Clearly the combination $\frac{\phi_{bg} - \phi}{\phi^\prime}$ is independent of the constant amplitude $C$ (which is what we want to find). So at a radius large enough that we are confident that the asymptotic solution applies, we may find the correct numerical solution by tuning $\phi_*$ such that $\frac{\phi_{bg} - \phi}{\phi^\prime}$ gives the correct value. This then uniquely determines the amplitude and thus gives out far field solution even for complicated and highly structured objects. The main barrier to this approach is the numerical stability of the solutions, which may be difficult to model for strongly perturbed fields. We shall not consider strongly perturbing fields further for this reason. 

\subsection{Fifth Force Interaction Between Two Objects}

The force on an object in general is the rate of change of momentum of that object $\vec{F} = \dot {\vec{p}}$. The derivation of this fifth force in the case of a Chameleon field around a small uniform sphere was given in Appendix B of \cite{Chameleon}. We shall therefore only highlight the key aspects, and note where our case deviates from theirs. In fact, the derivation is largely applicable with only a small modification to the force term. Two spherical objects at large separation, $A$ and $B$, are considered and the force on $B$ due to $A$ is calculated.

The Klein-Gordon equation for the scalar field is, in general, non-linear. This means that many body solutions are not generally superpositions of single body solutions. It is possible to superpose solutions in the case where the linearisation of the field equations is valid i.e. when the field is written $\phi = \phi_0 + \varphi$. In this case $\varphi_{A,B}$ individually satisfy the linear equations and thus \begin{equation} \phi = \phi_0 + \varphi_A + \varphi_B \end{equation} is a solution to the field equations. (Note that the field \emph{perturbations} are added, not the entire field solutions.) In this case, the form of the scalar field is simple to calculate, and the contributions from each object are well separated.

Inside the body we will assume that the scalar field is dominated by the body itself (as it is sensitive to the local density) so that the coupling term $C_B$ (which depends on the interior behaviour of the field in object B) is not sensitive to body A and vice versa. This allows us to use the scalar field solutions in equation \ref{spherical_field_solutions} for the field perturbations $\varphi_A$ and $\varphi_B$, centred around their respective objects. 

Expressing the total momentum of object B as an integral over a spherical volume around object B and differentiation, the force may be expressed as a surface integral around the spherical object, 
\begin{equation}
F_i = \dot P_i = -\int_V \partial_j \tau^i_j d^3x = -\int_S \tau_i^j n_j dS
\end{equation}
where $i,j$ run over spatial indices, and $\tau_{\mu\nu}$ is the energy momentum tensor for matter, the scalar field, and gravity (second order parts of the metric fluctuations). We use the notation $\tau_{\mu,\nu}$ for these energy momentum contributions for consistency with the derivation in \cite{Chameleon}. The gravitational part of the theory is unchanged in the Einstein frame and thus returns our usual Newtonian force law. The surface is drawn just outside of a bound object, so the matter terms are negligible here. The fifth force must come from the energy-momentum of the scalar field, which is the only part of $\tau_{\mu\nu}$ which does not appear standard GR case. To calculate the scalar field contribution to $\vec F$, we recall the energy momentum of the scalar field is
\begin{equation}
T^{(\phi)}_{\mu\nu} = \partial_\mu \phi \partial_\nu \phi - g_{\mu\nu} \left[ \frac{1}{2}(\nabla_\sigma\phi)(\nabla^\sigma\phi) -V(\phi) \right]
\end{equation}
and the gradient of our scalar field solution:
\begin{equation}\label{gradient_of_fields}
\nabla_i \varphi = \begin{cases}  \frac{Ce^{-m_{bg}r}}{m_{bg}r} \left[ r^{-1} + m \right] \frac{x_i}{r} 
			\\ 			\frac{C}{m_{bg}} \frac{x_i}{r^3} &\quad  r \ll m_{bg}^{-1}
			\end{cases}
\end{equation}
Since the surface is drawn around object $B$, and the coordinates are centred on $B$, we may use the latter approximation for $\nabla_i \varphi_B$ provided the radius of the object is much less than the characteristic scale of the force (which should be the case). The full derivative will need to be used later for gradient of the scalar field originating from $A$ as the distance from $A$ to $B$ is much larger than the radii of the objects themselves. Due to the geometry of the setup most of the contributions cancel out when integrated over the sphere, and only one term remains after the integration. We have:
\begin{equation}
F_i = -4\pi \frac{C_B}{m_{bg}} \partial_i \varphi_A
\end{equation}
Now switching our coordinate system and letting $r$ be the radial separation between objects $A$ and $B$, and inserting the gradient of the field $\varphi_A$ from equation \ref{gradient_of_fields}, we find the radial force law:
\begin{equation}\label{scalar_force}
F_\phi = -4\pi C_A C_B \frac{e^{-m_{bg} r}}{m_{bg}r} \left( 1 + \frac{1}{m_{bg} r} \right)
\end{equation}

The functional form of this is largely as one would expect from heuristic considerations (see \cite{Mota}), and involves a Yukawa force with a coupling to each body depending on the degree of screening which is exhibited by each. This attractive force can act as the source of scale dependent modified growth. 

Although \cite{Chameleon} derives a force law which is $r^{-2}$, and thus at large distances is indistinguishable from an additional mass, we are forced to confront the exponential decay of our fifth force. In the case of a dark matter free theory, this can lead to a discrepancy in the effective mass at short and large distances. MOND theories, which seek to stabilise rotation curves, are known to lead to low mass estimates inconsistent with the luminous mass of the galaxies when applied in the TA \cite{Zhao}. The exponential suppression in the far field force could help to remedy this discrepancy by requiring a larger mass to produce the same far-field acceleration. On the other hand, if the exponential suppression is too large, then one may find that the fifth force interaction which is making up for the lack of dark matter inside the galaxy become negligible on super-galactic scales. This would lead to approximately Newtonian dynamics, which produces a much higher mass estimate, suggesting a `missing mass' problem. 

\subsection{$M_{LG}$ in coupled theories and assumed properties of the Local Group}

Without selecting a particular theory, we may already use this generic form to investigate the effect on the LG, as is shown in Fig \ref{LG_ST_CM}. This will apply to any scalar field theory which generates these couplings and background effective masses. The $\Lambda$CDM limit is as $C_AC_B \rightarrow 0$ or $m_{bg} \rightarrow \infty$. The previous derivation of the fifth force does not include the acceleration produced by the energy density of the scalar field between the two objects (analogous to the scalar field, perfect fluid, and $\Lambda$ in the previous sections). In general this potential must be tuned to produce a $\Lambda$ like effect, and so we also explore the acceleration equation using a coupled scalar field with a cosmological constant $\Lambda$. 

\begin{figure} 
  \caption{Mass estimate contours for the LG using a generic spherically symmetric scalar field solution and varying the coupling and background effective mass of the scalar field.}
   \label{LG_ST_CM}
\includegraphics[width=8cm,keepaspectratio=true]{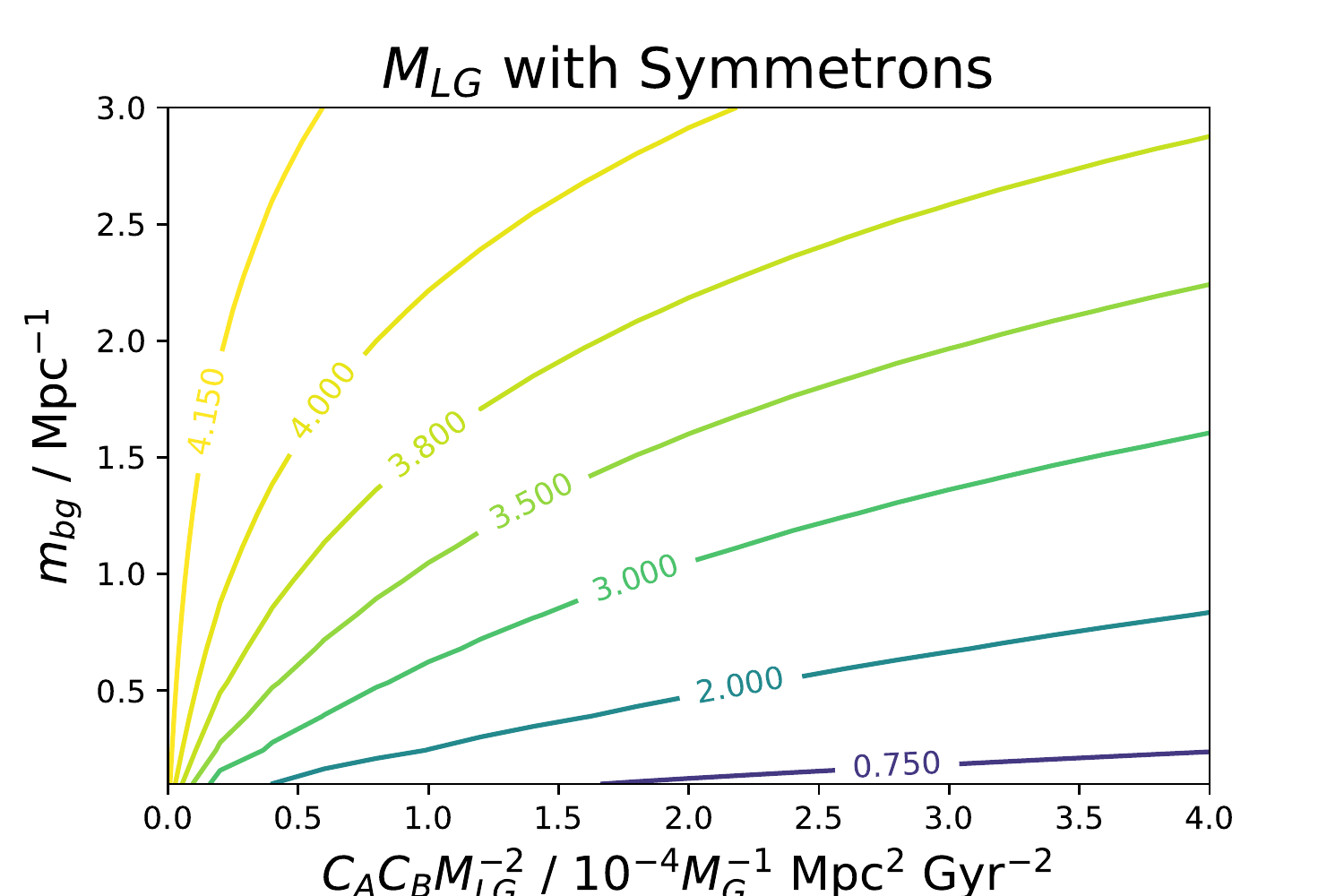}
\includegraphics[width=8cm,keepaspectratio=true]{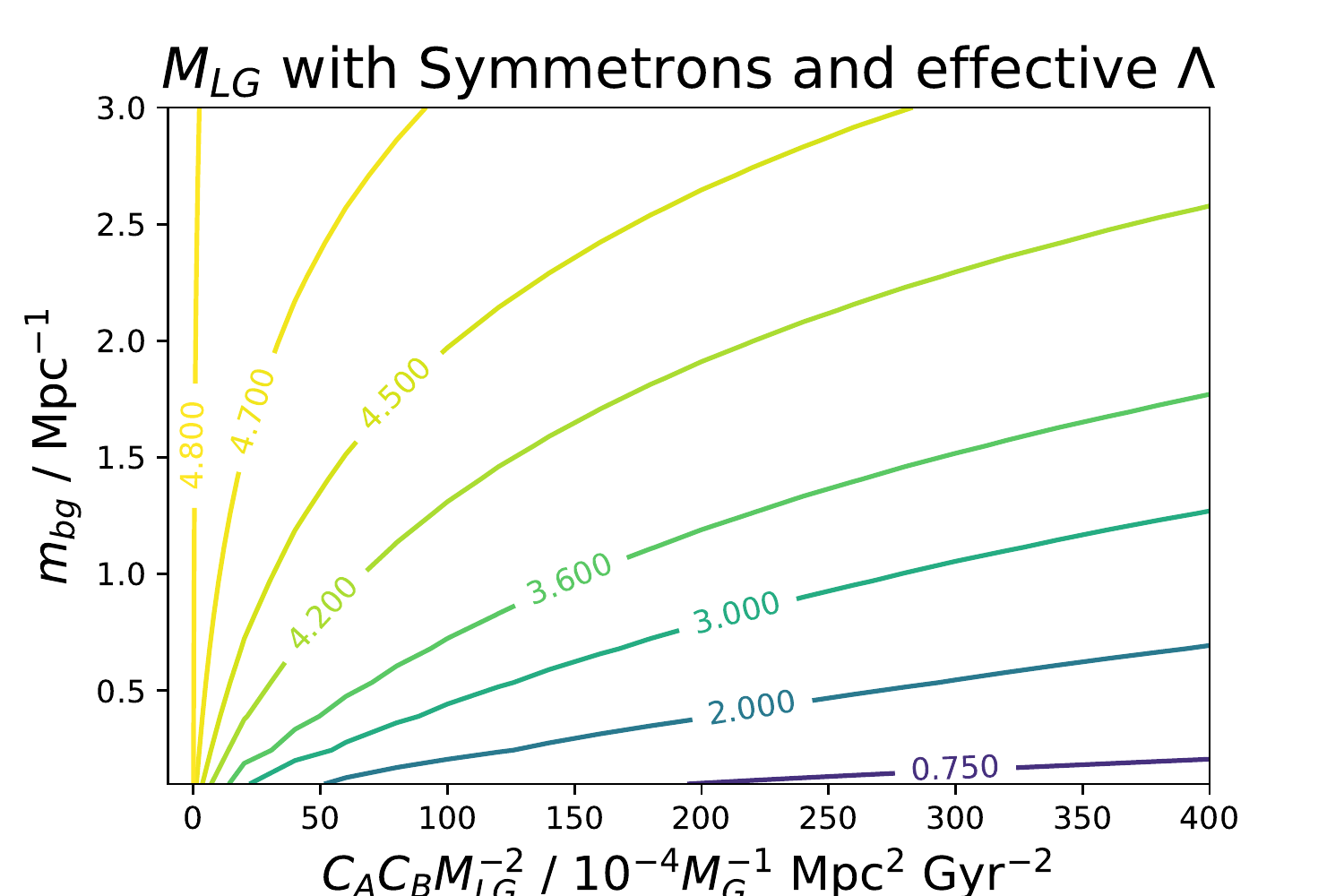}
\end{figure}

\subsection{A Dark Matter Free Symmetron Model}

Specific models may be explored by calculating the relevant scalar field effective mass $m_{\phi_{bg}}$, the background scalar field $\phi_{bg}$ and using the appropriate coupling for the theory. We consider a symmetron model here, based on the parameters in \cite{Symmetron}. In this case, the symmetron was introduced to stabilise rotation curves without the need for dark matter. Although this was based on numerical solution for observed galaxy profiles in cylindrical models, for the sake of being able to calculate a force law we will restrict ourselves for now to uniform spheres. 

The symmetron model has a quartic potential, of the form:
\begin{equation}
V(\phi) = V_0 - \frac{1}{2}\mu^2 \phi^2 + \frac{1}{4}\lambda \phi^4 
\end{equation}
where $\mu^2, \lambda > 0$. The effective potential is coupled to the matter distribution, although far from the object the matter distribution is negligible. The minimum of the potential is at 
\begin{equation}\label{phi_bg}
\phi_{bg} = \pm \frac{\mu}{\sqrt{\lambda}}
\end{equation}
The effective mass at the minimum, $m_{\phi_{bg}}$, is given by:
\begin{equation}
m_{\phi_{bg}} = V^{\prime \prime}(\phi_{bg}) = 2\mu^2 
\end{equation}
The final ingredient we require is the matter coupling, which is given by 
\begin{equation}
A(\phi) \approx 1 + \left(\frac{\phi}{M}\right)^2 + \mathcal{O}\left[ \left(\frac{\phi}{M}\right)^3\right]
\end{equation}
As an aside, it is useful to look at the potential of such a scalar field background:
\begin{equation}
V(\phi_{bg}) = V_{0} - \frac{\mu^2}{4\lambda}
\end{equation}
If the gradient of the scalar field is negligible compared to the potential (likely to be true over areas where the density is roughly uniform) then this means the energy density of the scalar field $T^\phi_{\mu\nu}$ is dominated by the potential, which is always $< V_0$ by definition. This means that in order to produce a dark energy effect, symmetrons would be required to have a positive definite $V_0$, which is, mathematically at least, equivalent to adding a cosmological constant into the theory. We will thus also consider the behaviour of a symmetron model with $\Lambda$. 

The more general case in the presence of a matter density $\rho_m$ is required to understand the contribution of the scalar field to gravitation. The crucial properties are the ground state of the scalar field $\phi_g$ and the effective mass there $m_g$. 

\begin{equation}
V(\phi,\rho) = V_0 + \frac{1}{2}(\mu^2_\rho - \mu^2)\phi^2 + \frac{1}{4}\lambda \phi^4
\end{equation}
where $\mu_\rho^2 = \frac{\rho}{M^2}$. Note that when $\mu_\rho > \mu$ (high density environments) then the potential has only one stationary point, a minimum at $\phi = 0$. At lower density, the central stationary point become a local maximum, and there are two minima at 
\begin{equation}\label{phi_c}
\phi_0 = \pm \sqrt{\frac{\mu^2 - \mu_\rho^2}{\lambda}}
\end{equation}
which tends to the background solution (Eqn. \ref{phi_bg}) for $\rho_m = 0$. 
The mass is given by:
\begin{equation}\label{m_c}
m^2 = 2(\mu^2 - \mu_\rho^2)
\end{equation}
In the case where $\phi_0 = 0$ ($\mu_\rho > \mu$), then the mass is $\mu^2_\rho - \mu^2$. 

Using the expression for the coupling (Eqn. \ref{coupling}), we can see that as the density of the object increases, the coupling will decrease. (Although $\Delta\phi$ will increase, the $[x\coth(x)-1]^{-1}$ term grows more quickly.) This leads to very dense objects decoupling from symmetron fifth-force effects. 

The free variables here are then $\mu$, $\lambda$, and $S$; we may take $\Omega_\Lambda$ and the background expansion to be typical of $\Lambda$CDM. One may also wish to vary the structural properties of the galaxies which are not necessarily well observed, such as the effective radius of the halo, but we shall not do this here. 

Screening is an important part of modified gravity models. Contrary to many models, the effective mass of the scalar field is actually lower in high density environments, so that the force becomes very long ranged. Screening in the symmetron model is instead achieved by having the scalar field value drop to zero in high density environments. This reduces the conformal coupling $A(\phi) \rightarrow 1$, which returns GR. (There may remain a vestigial potential $V(\phi = 0) = V_0$ which would act as a cosmological constant.) When the local density is low enough the minimum of the potential rises so that $|\phi_0| > 0$, and the effects of the field may become apparent. In order for the symmetron to have a significant effect on the motions of the galaxies, we require the field to be poorly screened on the scale of the galaxy. It can nevertheless be well screened in even denser environments, such as the solar system. 

\begin{figure} 
  \caption{Scalar field solution inside and outside of a sphere of radius 30kpc, and mass $10^{12}M_\odot$. At the object boundary and outwards, the scalar field is very close to the background value, which shows that we can treat this as a perturbation to the background value. The scalar field drops inside the object, but remains high enough to show that the object is not screened internally. $\phi_{bg}$ is the ground-state of the field in the background (external to the object), as in equation \ref{spherical_field_solutions}.}
   \label{ScalarSolution}
\includegraphics[width=12cm,keepaspectratio=true]{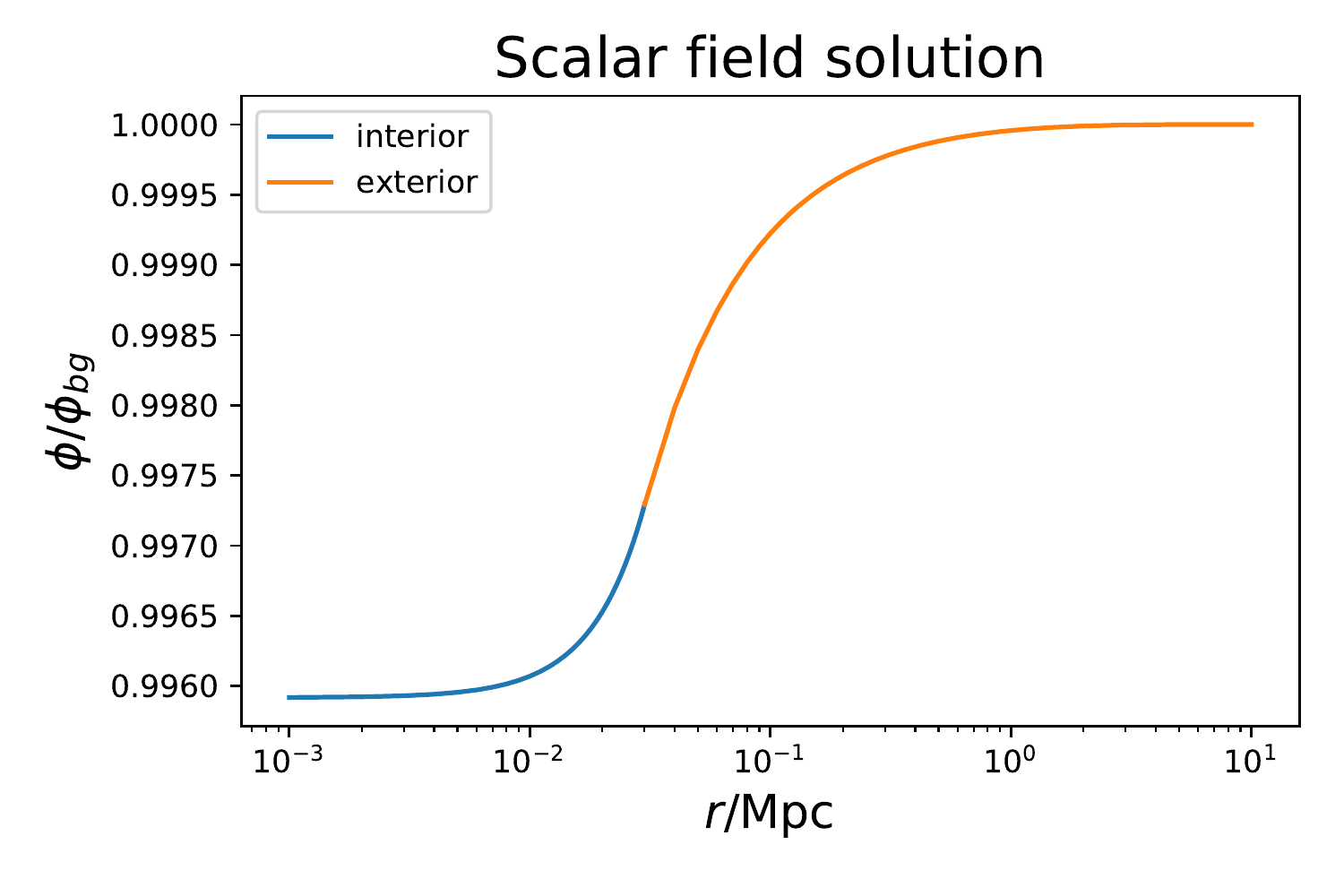}
\end{figure}

An example of the scalar field solution inside and outside of a spherical object is given in Fig \ref{ScalarSolution}. The internal solution shows that, as expected, the interior of the object is not screened, which allows the scalar field to have an impact on the rotation curve. The external solution shows the exponential decay of the perturbation to the scalar field, and shows that the perturbation to the background value is small, allowing us to make use of the approximations in the previous sections. 

\begin{figure} 
  \caption{Mass estimate contours for the LG using a symmetron model and varying the free parameters in the potential function. The MW and M31 have been taken to be uniform spheres with radii 30 and 33 kpc respectively, with $M_{M31} = 2 M_{MW}$. The black dot corresponds the the fiducial values in \cite{Symmetron}. The mass estimate is significantly lower than in GR due to the attractive scalar field fifth force; nevertheless it is still nearly an order of magnitude too large to be accounted for by the baryonic mass of the two galaxies.}
   \label{LG_ST_SY}
\includegraphics[width=8cm,keepaspectratio=true]{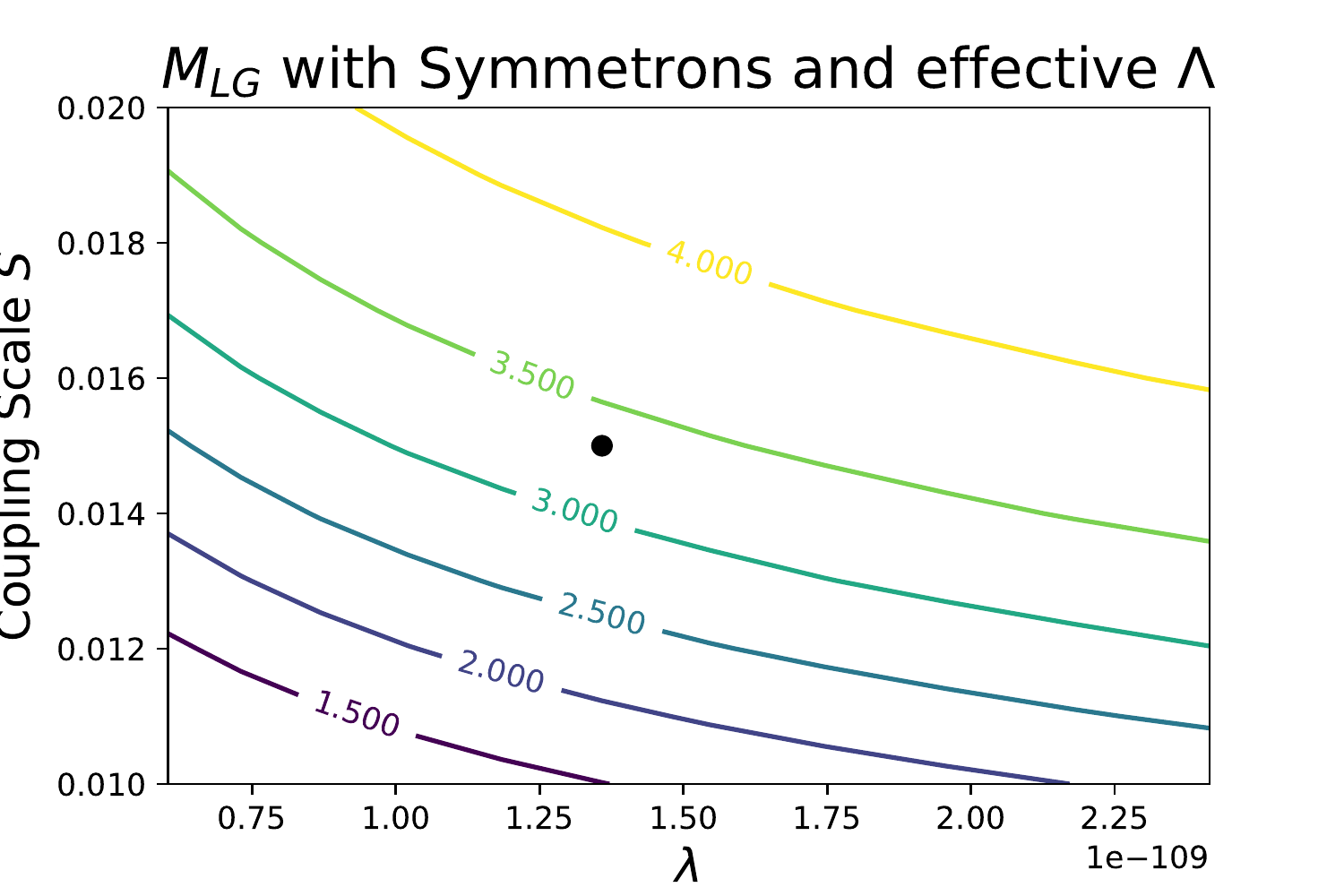}
\includegraphics[width=8cm,keepaspectratio=true]{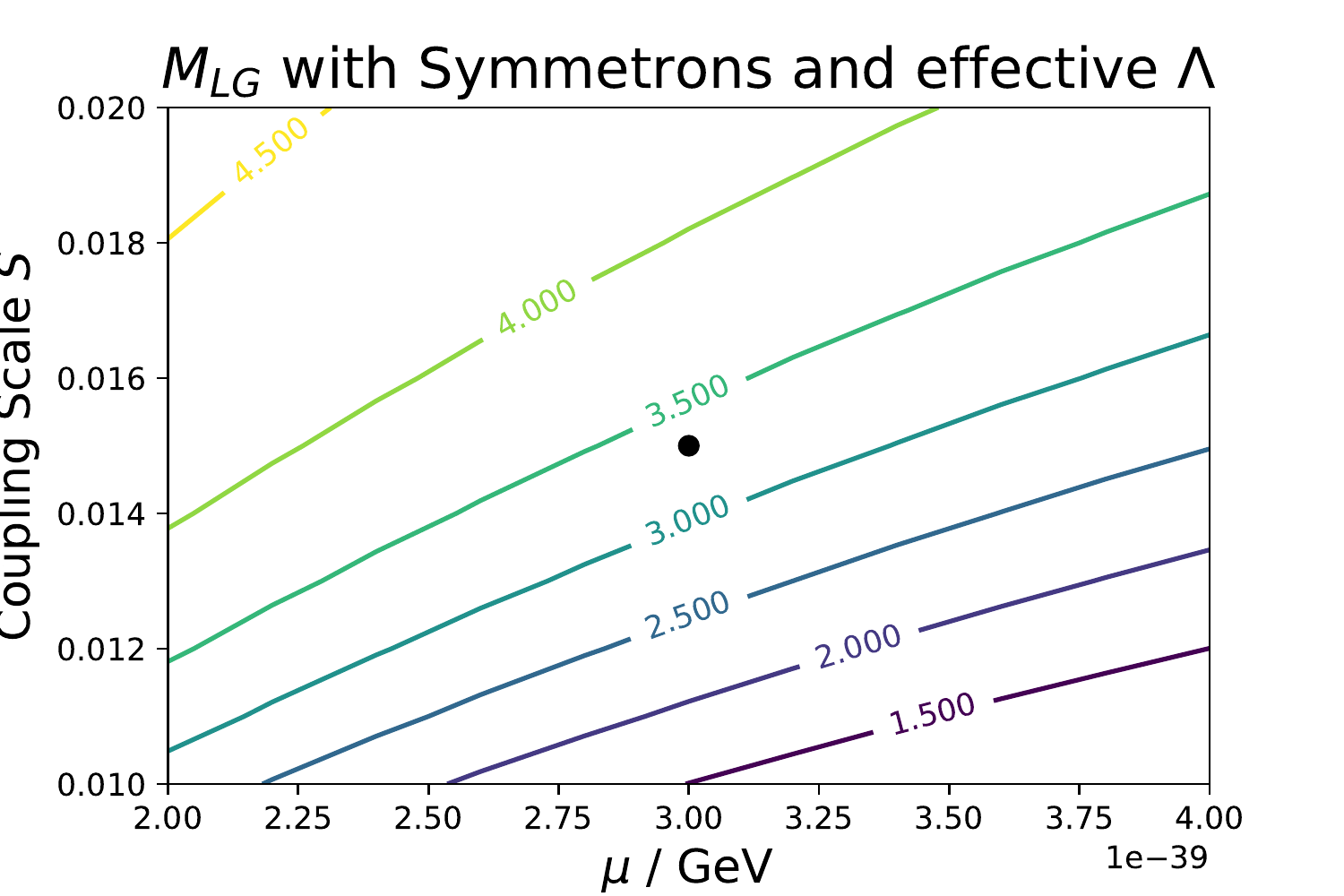}
\end{figure}

Using the symmetron model we can explore the effects of the symmetron potential parameters directly, as in Fig \ref{LG_ST_SY}. We vary our parameters close to the values used in \cite{Symmetron} for stabilising galactic discs. Notice that the estimated mass for the values given in \cite{Symmetron} (shown as the black dots on Fig \ref{LG_ST_SY}) is much larger than the luminous mass of MW and M31, which could indicate a significant inconsistency in using this theory of gravity to eliminate dark matter. It should be noted that more detailed modelling would need to be used to confirm this, such modelling the galaxies as disks with realistic density profiles, and abandoning spherical symmetry in favour of cylindrical. (This leads to significant complications if the disks are not co-planar.) Furthermore, mass external to the galactic disc such as satellites should be modelled for their contribution to the scalar field solution. Despite the shortcomings of our model, as a first approximation it suggests a potential missing mass problem in this model.

\begin{figure} 
  \caption{Mass estimate contours for the LG using a symmetron model and varying the radii and mass ratio of the galaxies. On the left we vary the radii of the MW and M31 between 20 and 40 kpc, which has only a slight impact on the mass estimate. On the right we vary the radius of the MW and the fraction of the mass in the MW: $f_{MW} = \frac{M_{MW}}{M_{MW}+M_{31}}$. Black dots are shown at the fiducial parameters $r_{MW} = 0.030$Mpc, $r_{M31} = 0.033$Mpc, and $f_{MW} = \frac{1}{3}$. Symmetron field parameters are kept fixed at their fiducial parameters.}
   \label{LG_ST_RF}
\includegraphics[width=8cm,keepaspectratio=true]{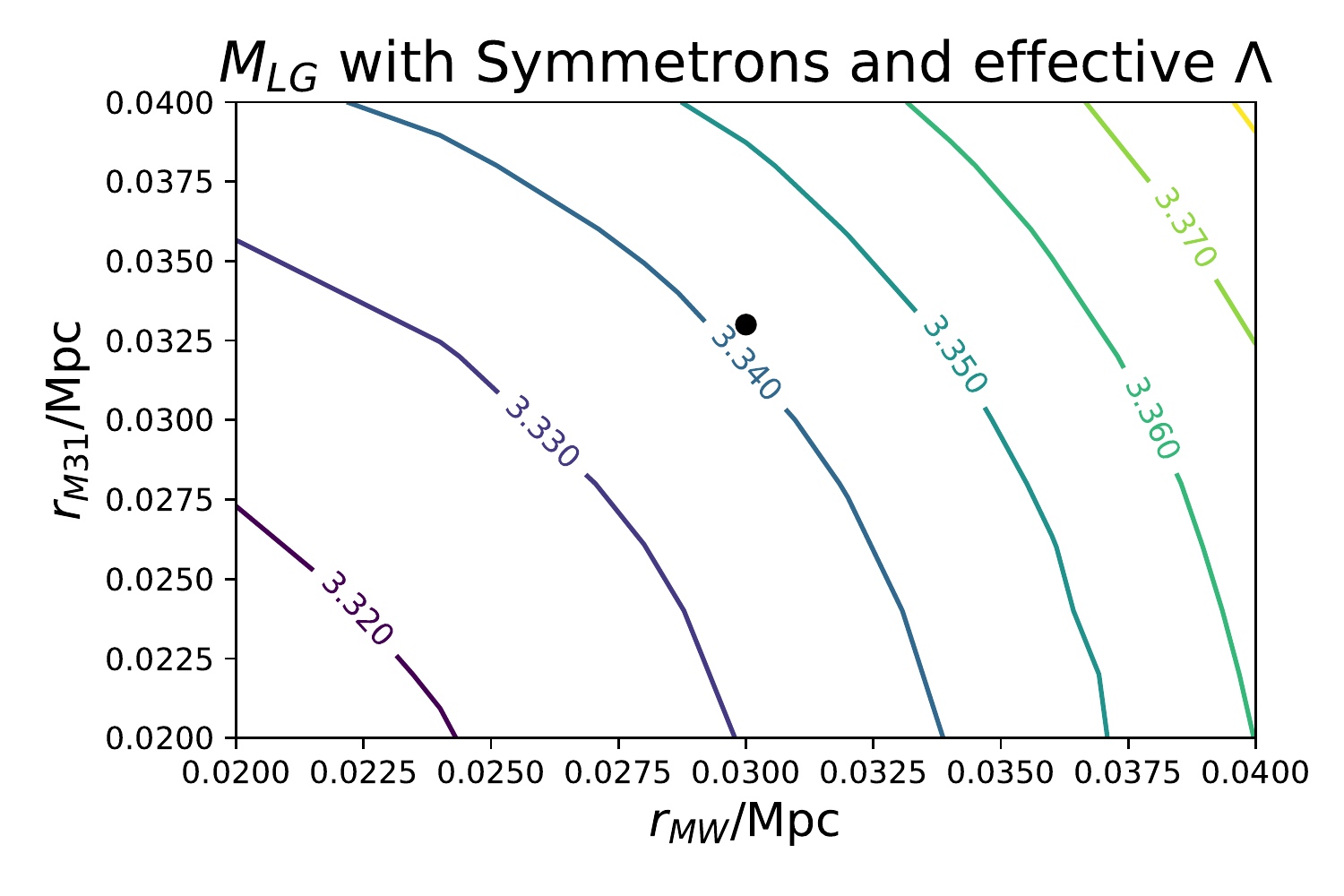}
\includegraphics[width=8cm,keepaspectratio=true]{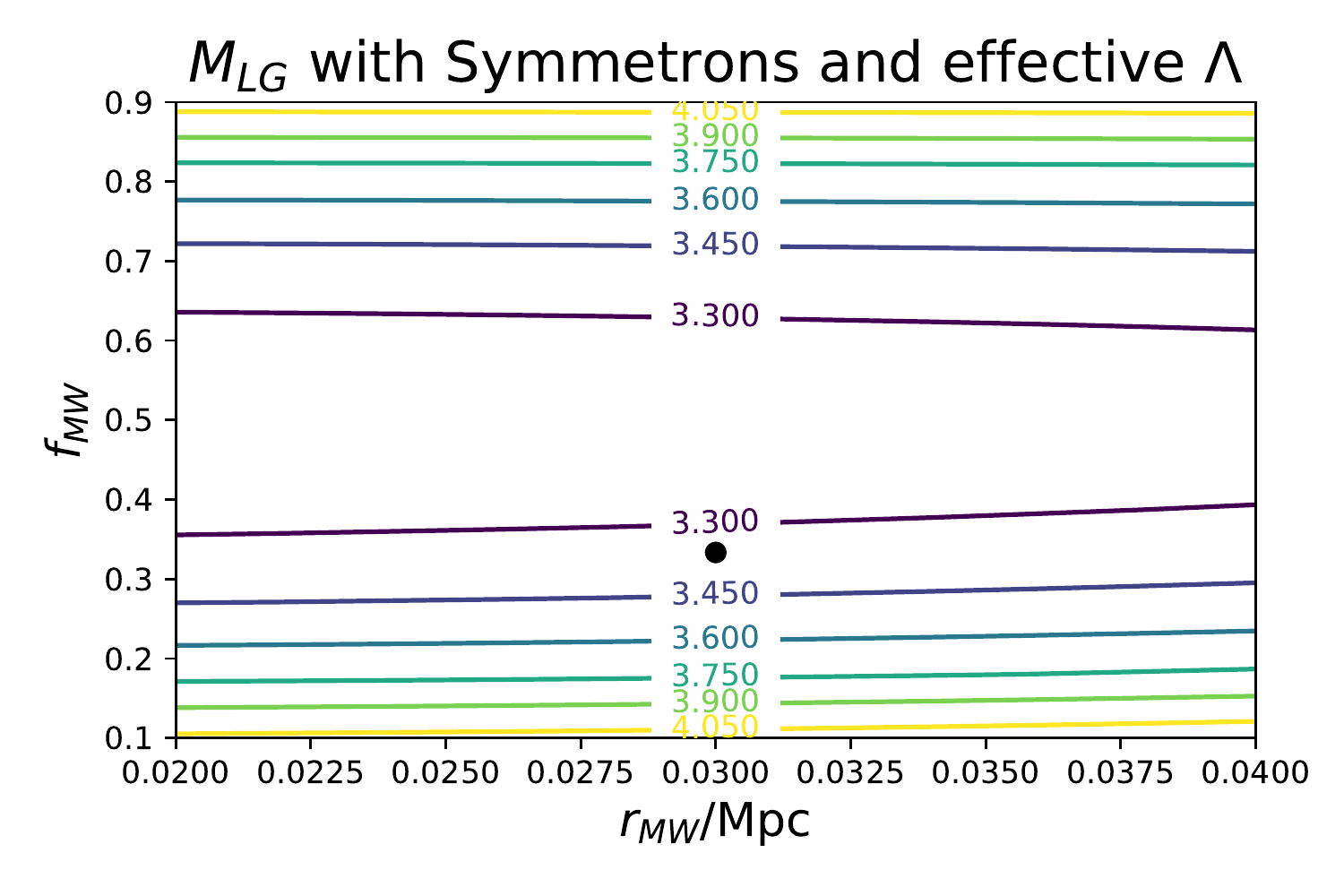}
\end{figure}

We can also explore how the radii and mass ratio of the two galaxies affect the estimate of the mass using this model. In Fig \ref{LG_ST_RF} shows that the mass estimate is not sensitive to the radii of the spheres compared to the uncertainties in other parametrs, but the mass ratio can have an impact of $\mathcal{O}(10^{11}M_\odot - 10^{12} M_\odot)$ for mass distributions where the vast majority of the mass is bound in one of the galaxies. Such a distribution of mass is not realistic however, and the mass estimate cannot be brought down beyond $3\times10^{12}M_\odot$ by a change in the mass ratio, which means that uncertainty in this parameter cannot account for the excess mass without dark matter. 

\section{Modified Newtonian Dynamics (MOND)}\label{TA_MOND}

MOND is another alternative to dark matter, which was first proposed in 1983 \cite{Milgrom}. Since its inception a large number of variants have been produced which produce the same limiting case dynamics in different ways. Its best known achievements are fitting galactic rotation curves and providing an explanation for the Tully-Fisher relation. Nevertheless, a gravitational law which stabilises a galactic rotation curve does not necessarily have desirable long range behaviour; the luminous mass of the MW and M31 must be able to generate a similar far field acceleration to dark matter in order to produce the observed dynamics. One might immediately suspect that this could be an issue, since in the deep-MOND regime (valid where the galaxies are well separated and the gravitational interaction is very weak -- much weaker indeed than the gravity at the edge of the disk) one has $a \approx \sqrt{a_N a_0}$ i.e. the geometric mean of the Newtonian acceleration and the fixed acceleration scale $a_0$. This then naturally produces $a \sim \sqrt{r^{-2}a_0} \propto r^{-1}$. We therefore expect the long range gravitational acceleration to be excessive, and thus the $\text{TA}_\text{MOND}$ mass estimate to be too small. This is consistent with \cite{Zhao} for example, who find that MOND implies a close encounter in the past between MW and M31. Since the model used throughout this paper is purely radial (and thus is doomed to produce singular acceleration in any scenario where MW and M31 pass) we shall not model such an event, although we discuss some of the implications below. 

The results in MOND will depend upon the universal acceleration scale, which can be independently determined from galaxy rotation curves. Calculating a consistent cosmological history in a non-GR theory such as MOND is non-trivial. We are not in a position to calculate the age of the universe in a fully relativistic formulation of MOND, although we can test the MOND model using a sensible range of ages for the universe $11-15$ GYr. Although such trivial formulations of MOND are not consistent with cosmology \cite{HomoMOND}, we may explore the effects of MOND in a local way under the assumption that there exists a theory which does not violate cosmological principles on large scales and reduces to MOND in the appropriate limit. 

\subsection{Acceleration Equations in MOND}

The non-relativistic MOND theory has a non-linear modified Poisson equation \cite{MOND_Review}:
\begin{equation}
\nabla \cdot \left[ \mu\left(\frac{a}{a_0} \right) \nabla \Phi \right] = 4 \pi G \rho = \nabla \cdot \nabla \Phi_N
\end{equation}
From this we may infer:
\begin{equation}
\mu\left(\frac{a}{a_0} \right) \nabla \Phi = \nabla \Phi_N + \nabla \wedge F
\end{equation}
where the curl field $\nabla \wedge F$ is taken to be zero to match the boundary conditions that $\nabla\wedge F = 0$ at $r \rightarrow \infty$ and immediately around a spherical object (since the acceleration field should be radial), as well as keeping the correct limiting behaviour for the Newtonian and deep-MOND regimes. The `standard' MOND term uses the parameterisation
\begin{equation}
\mu \left(\frac{a}{a_0}\right) = \left[ 1 + \left( \frac{a_0}{a} \right)^2 \right]^{-\frac{1}{2}} 
\end{equation}
Rewriting $-\frac{GM}{r^2} = a_N$ we arrive at the acceleration equation 
\begin{equation}
a^4 = a_N^2 a^2 + a_N^2 a_0^2
\end{equation}
which is just a simple quadratic equation. Given that the determinant $a_N^4 + 4a_N^2a_0^2 > a_N^4$ there is one positive solution for $a^2$. We finally have 
\begin{equation}
a = -\left[ \frac{ a_N^2 + \sqrt{ a_N^4 + 4 a_N^2 a_0^2 } }{2} \right]^{\frac{1}{2}}
\end{equation}
for the standard interpolating function. 

There is also a `simple' interpolating function, which leads to a slightly different solution. We will investigate the effects of both. The simple interpolating function is 
\begin{equation}
\mu\left(\frac{a}{a_0}\right) = \left[ 1 + \frac{a_0}{|a|} \right]^{-1}
\end{equation}
which leads to a closely related acceleration equation:
\begin{equation}
a = - \left[ \frac{ |a_N| + \sqrt{ |a_N|^2 + 4 |a_N||a_0| }}{2} \right]
\end{equation}
We may investigate the mass of the LG when varying $a_0$ and $t_u$. (We do not vary the cosmological parameters directly in this case as it is not clear what the cosmological implications of non-relativistic MOND is.) 

In the case of a LG analysis, both interpolating functions give similar results as we are typically in the `deep MOND' regime, and only transition to high accelerations fleetingly at very close distances. The results of a MOND based analysis are masses which are approximately 1\% of their usual Newtonian estimates. This presents significant tension with observational results, which suggest that the usual baryonic mass of the MW and M31 significantly exceed this (making up approximately 10\% of the Newtonian estimate). 

A possible solution as presented in \cite{Zhao} is to permit a close encounter in the past, so that the galaxy pair is now on their second pass (which of course requires non-radial motion). It has further been suggested in \cite{Bilek} that, though a close encounter is likely to be unfeasible in $\Lambda$CDM without more disruptive effects and a likely merger, the galaxies may survive with some structures surrounding the MW and M31 naturally explained by MOND simulations involving a close pass. In \cite{Banik_1}\cite{Banik_2} MOND simulations with a fly-by event are also used to explain the planar structure of satellites around MW and M31. Reliable exploration of MOND in this direction requires a more detailed model which incorporates both the angular momentum of the LG (to avoid a singularity at $r=0$) and the external gravitational field (since MOND is highly non-linear), and is beyond the scope of this paper. The MOND models with a past encounter reflect a LG mass of $2-3 \times 10^{11} M_\odot$, consistent with expectations. For further analysis of past encounters see \cite{Benisty} (in preparation). 

\begin{figure} 
  \caption{Mass estimate contours for the LG using MOND with the simple and standard interpolation functions. The masses are significantly lower than the luminous masses of MW and M31, by approximately an order of magnitude. This is consistent with who find that for MOND to be consistent with the TA, the galaxies need to be on their second pass.}
   \label{LG_ST_CM}
\includegraphics[width=8cm,keepaspectratio=true]{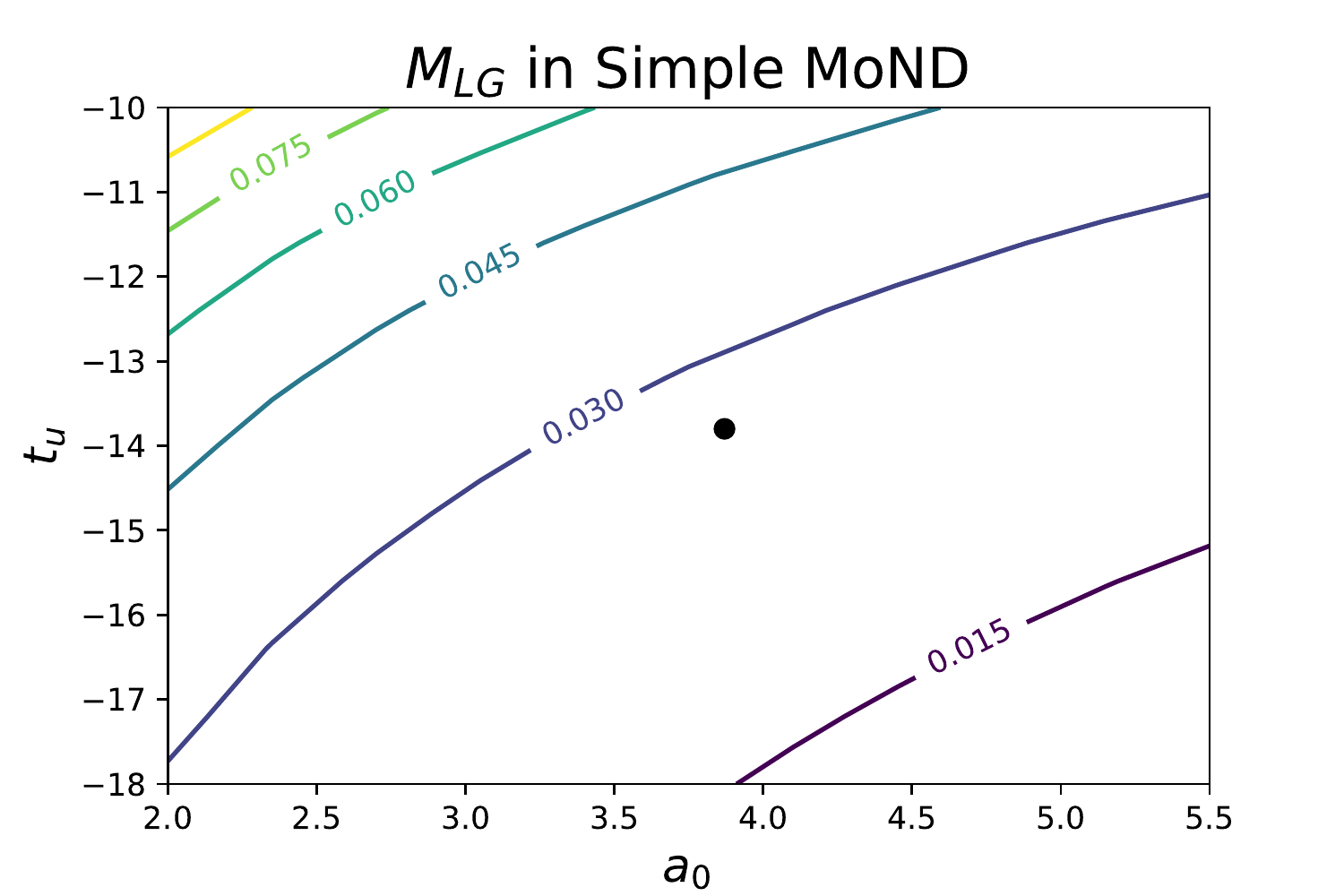}
\includegraphics[width=8cm,keepaspectratio=true]{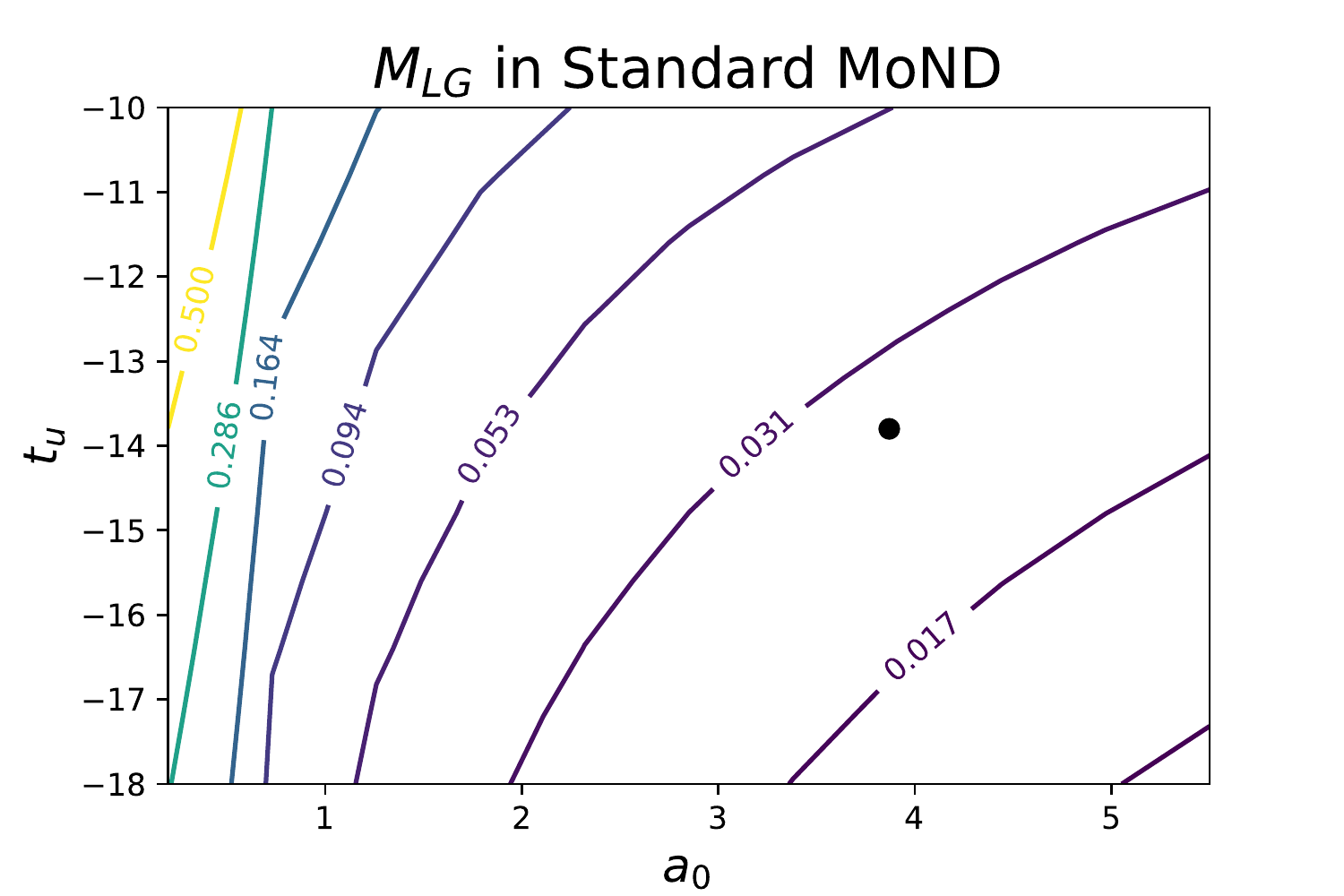}
\end{figure}

\section{Discussion}

By considering the LG as a two body problem, we estimate its mass for a variety of models and parameters. We summarise some examples of LG mass estimates in table \ref{Estimates_Summary} for a variety of models and parameters. 

When looking at the impact of cosmology upon the LG mass estimate, different models of dark energy in models with a significant cold dark matter component produce relatively weak effects on the LG mass estimate, although values of $h$ at the extremes of the observed range produce changes in $M_{LG}$ which are of the same order of magnitude as uncertainties in observables such as radial separation, velocity, and the uncertainty due to the considerable simplifications in the model itself. This means that we cannot ignore cosmology when calculating the LG mass, and we should always consider these estimates in their proper context. Whilst the effects on the LG mass are not extreme enough to produce masses which we can claim are inconsistent with other observations on the LG -- and thus we cannot turn this analysis around and place a limit on $\Lambda$CDM -- it appears that cosmology could be an important component of precise estimates of the LG mass in future. 

When looking at the inferred LG mass in our two dark matter free theories -- MOND and symmetrons, designed to explain flat rotation curves and emphasising galactic scales -- we find mass estimates which appear discrepant with the luminous mass of the Milky Way and Andromeda. In MOND we find a mass which is too small to account for the observed matter in these galaxies; this is usually explained by the possibility of a previous encounter between the MW and M31 \cite{Zhao}, which cannot be explored in this one-dimensional TA due to the pole in the acceleration equation at $r=0$. In the case of the symmetron model that we explored, although the modelling is highly idealised, the estimate of the mass is an order of magnitude too large to be consistent with a dark matter free mass estimate, suggesting that there is a missing mass problem in this model. (Unlike MOND, this approach cannot be salvaged by assuming a previous encounter, since the mass estimate is high rather than low.) Detailed numerical modelling of the extended structure, including modelling of the galactic discs and their satellites, are beyond the scope of this paper, but would be useful in assessing the viability of this model on Mpc scales. 

In principle, the LG and other similar Mpc scale systems could be used to make inferences about gravity and dark energy, by reversing the approach we have so far taken. If, by some independent means (such as strong gravitational lensing or stellar streams), one \emph{knew} the mass of the system (in the context of a given model), one could then constrain gravity by looking at the mass estimates of the Local Group. We do this to some extent when we look at MOND or symmetrons. In this case, the mass of the galaxies is understood reasonably well by estimating the luminous mass of the Milky Way and Andromeda; in order to be viable, one must find a MOND or symmetron model which provides a consistent mass estimate. Stellar streams may also be used to estimate the mass of the Milky Way through gravitational effects, but on a much smaller scale. The consistency between these scales would be the key to understanding the gravitational theory. We can see from the contour plots in the previous sections that each of these theories has free parameters which are degenerate with respect to the LG mass estimate; in order to constrain the theory more fully one would have to break this degeneracy with other observations. Although the use of the LG as a local laboratory for dark energy remains intangible for now, the behaviour of the LG still provides a valuable consistency check on proposed theories of gravitation at these often neglected scales. As we have already seen, a modified gravity theory which binds a galaxy together to produce accurate rotation curves without dark matter can have problems reproducing longer range effects such as galaxy-galaxy interactions because of the different scaling of the effective gravitational force law compared to Newtonian gravity.  

\begin{table}
\centering
\caption{Examples of LG mass estimates for different models and parameters. Parameters in the symmetron model are given here in Planck units, and are based on \cite{Symmetron}. The variation between models can be quite large, even when remaining in $\Lambda$CDM due to the tension over $h$. Quintessence has not been included in the table due to a lack of available direct constraints over an exponential potential. $\Lambda$CDM provides a limiting case for quintessence models, but without observational constraints there are no exemplary values to use in assessing the impact of the quintessence field on the LG mass. The variation of the mass given different values of the quintessence parameters is however shown in figure \ref{LG_MC}.}
\label{Estimates_Summary}
\begin{tabular}{|c|c|c|} \hline
Model & Parameters & $M_{LG}$ / $10^{12} M_\odot$ \\ \hline 
$\Lambda$CDM & $h = 0.67$, $\Omega_\Lambda = 0.7$ & 4.80 \\ \hline
$\Lambda$CDM & $h = 0.63$, $\Omega_\Lambda = 0.7$ & 4.55 \\ \hline
$\Lambda$CDM & $h = 0.76$, $\Omega_\Lambda = 0.7$ & 5.43 \\ \hline
$w$CDM & $h = 0.67$, $\Omega_f = 0.7$, $w=-1.1$ & 4.79 \\ \hline
$w$CDM & $h = 0.67$, $\Omega_f = 0.7$, $w = -0.9$ & 4.81 \\ \hline
Symmetron + $\Lambda$ & $\mu = 2.46\times 10^{-58}\us\text{GeV}$, $\lambda = 10^{-109}$, $S = 0.015$ & 3.34 \\ \hline
MOND & $a_0 = 1.2\times 10^{-10}\us\text{m}\us\text{s}^{-2}$, $t_u = 13.69\us\text{Gyr}$ & 0.027 \\ \hline
\end{tabular}
\end{table}

\acknowledgments

The authors would like to thank Ed Copeland for his invaluable discussions on scalar-tensor theories, and Mordehai Milgrom for his guidance on MOND. We also thank David Benisty, Eduardo Guendelman, Yehuda Hoffman and Noam Libeskind for stimulating discussions.
OL acknowledges support from a European Research Council Advanced Grant FP7/291329 and STFC Consolidated Grant ST/R000476/1.


\end{document}